\documentclass[%
 reprint,
superscriptaddress,
 amsmath,amssymb,
 aps,
]{revtex4-2}

\usepackage{graphicx}
\usepackage{dcolumn}
\usepackage{bm}
\usepackage{color}
\usepackage{hyperref}

\begin{document}

\preprint{APS/123-QED}

\title{Textural properties of dense granular pastes produced by kneading}

\author{Mathilde Auxois}
 \affiliation{IFP Energies nouvelles, Rond-point de l’échangeur de Solaize, BP 3, 69360 Solaize, France\looseness=-1}
 \affiliation{ENSL, CNRS, Laboratoire de physique, F-69342 Lyon, France}
\author{Marine Minière}
\author{Chloé Bertrand-Drira}
\author{Fabien Salvatori}
\author{Jan Verstraete}
 \affiliation{IFP Energies nouvelles, Rond-point de l’échangeur de Solaize, BP 3, 69360 Solaize, France\looseness=-1}
\author{Sébastien Manneville}
\author{Thibaut Divoux}
\affiliation{ENSL, CNRS, Laboratoire de physique, F-69342 Lyon, France}

\date{\today}

\begin{abstract}
The efficiency of supported catalysts depends on the porous microstructure of their solid support, which regulates mass transfer, exposure to the active phase, and mechanical strength. Here, we focus on the manufacturing of a specific type of catalytic support, $\gamma$-alumina extrudates, by a kneading-extrusion process. In this process, a paste is initially formed by blending and subsequently kneading a boehmite powder, an aluminum oxide hydroxide precursor of $\gamma$-alumina, with various liquids before undergoing extrusion. The crucial step in this process is the kneading step, which allows control over the textural and mechanical properties of the extrudate. The present experimental study aims to measure the impact of the kneading step on boehmite pastes prepared using a pilot kneader. The pastes are obtained by mixing boehmite powder with an acid and a basic solution in a two-step process known as peptization and neutralization. In this study, kneading is conducted at various mixing speeds and durations while monitoring the torque exerted by the boehmite paste on the kneader blades. Moreover, samples are extracted from the pilot kneader at various stages of the kneading process, and their textural properties are determined by both nitrogen sorption and mercury intrusion porosimetry. Our findings show that, at fixed composition, the textural properties of boehmite pastes are controlled by the overall deformation accumulated during kneading. Conversely, for a fixed accumulated deformation, the pH sets the textural properties. Finally, we identify an empirical control parameter that captures the combined effects of pH and accumulated deformation on the key textural attributes of boehmite pastes. These results set the stage for a systematic design approach of boehmite-based catalyst supports.
\end{abstract}

\maketitle


\section{Introduction}
\label{sec:Introduction}

Blending powders with liquids is a common process performed in different industrial sectors such as civil engineering, food production, and pharmaceuticals \cite{Coussot:2005,Harnby_1997,Goldszal:2001, Dhanalakshmi:2011, Hansuld:2014}. The textural characteristics of the mixture vary depending on the powder-to-liquid ratio. At lower liquid content, the mixture may exhibit properties of a cohesive powder or crumbly agglomerates, whereas, with an adequate amount of liquid acting as a continuous phase, it can transform into a dense, cohesive granular paste
\cite{Rondet:2008, Rondet:2009a}. 

Processes designed for preparing pastes involve handling highly viscous liquids, typically with viscosities higher than 10~Pa.s, or even solid-like viscoelastic materials \cite{Paul_2003}. Such processes are referred to as ``\textit{kneading}." The goal of kneading operations is to induce fluid movement through laminar stretching and folding so as to disperse agglomerates, favor chemical reactions, and eventually achieve sample homogenization \cite{Paul_2003,Campelli_2020,Adragna_2006}. Due to the substantial viscosity of the mixture resulting in significant dissipation, the fluid motion is primarily driven by the rotation of mixing blades throughout the volume. Effective technology design is crucial in this process, leading to the development of batch or ~continuous mixing technologies encompassing different tank and blade shapes, sizes, and complexities \cite{Paul_2003,LeLan_1983,Albright_2009,Seem:2015}. 

The present study focuses on the kneading process employed in the industrial production of $\gamma$-alumina extrudates, which are currently used as catalyst supports for hydrotreating in refineries   \cite{Lloyd_2011,Dubois:1995, Vatutina_2016}. These extrudate supports are shaped by a four-step kneading-extrusion process involving kneading, extrusion, drying, and calcination aiming at transforming boehmite powder --a $\gamma$-alumina precursor-- into cylindrical objects a few centimeters long \cite{Euzen_2008}. Industrially, the microstructure of catalytic supports plays a critical role in determining the quality of the carrier by controlling mass transfer and mechanical strength \cite{Klaewkla:2011,Hart_1990,Trimm_1986,Kolitcheff:2017}. Previous work on extrudate manufacturing has explored methods to control support structures throughout the shaping process in order to improve their mechanical robustness and fine-tune their textural properties. Specifically, earlier studies have separately examined the thermal treatment phase of the process (i.e., drying and calcination), from the kneading and extrusion stages \cite{Pourcel_2007,Trimm_1986,Karouia_13}. Among the key results of these studies, it has been demonstrated that the kneading step can impact the porosity of the catalytic support via two parameters: the support composition, and the type of kneading equipment used, along with its operational conditions. 

On the one hand, the support porosity can be modified by mixing boehmite powder with peptizing agents, typically nitric and boric acids, which induce a colloidal dispersion of the boehmite powder \cite{Speyer_2020,Ramsay:1978,Fauchadour_2002}. The peptization depends on factors like the molar ratio of the peptization agent to the boehmite powder, the nature of the peptizing agent (type of acids, ammonia), and the inherent dispersibility of the boehmite powder fixed during the boehmite synthesis \cite{Speyer_2020,Karouia_13,Zheng_2014,Dubois:1995,Vatutina_2016}. More specifically, it was shown that boehmite peptization reduces the population of macropores (i.e., pores with a diameter exceeding 50~nm) without significantly affecting the micropores and mesopores (i.e., pores with diameters smaller than 50~nm) \cite{Trimm_1986,Danner:1987}. 
Another approach to tuning the support porosity involves mixing boehmite powder with additives such as polymers (e.g., methylcellulose, polyethylene glycol) or organic species (e.g., ethylene glycol, glycerin) \cite{Trimm_1986}. Both types of additives are used to create macro- or meso-porosity depending on their size, acting in two ways: freeing space when removed by combustion or modifying the spatial arrangement of alumina precursor particles by physico-chemical interactions \cite{Trimm_1986,Studart:2006,MarquezAlvarez:2008}. 

On the other hand, the kneading technology used, including the shape and volume of the tank, the number, position, and geometry of the blades, and the operating conditions also impact the support porosity, although this aspect has received less attention in the literature \cite{Walendziewski:1994,Landers_2010}. Specifically, for boehmite powder, increasing the kneading duration impacts the bimodal pore size distribution of the alumina support, promoting the growth of a larger pore population (with a mean diameter around 13~nm), while reducing the smaller pore population (with a mean diameter around 8~nm) \cite{Landers_2010,Dubois:1995}. 
However, the majority of previous studies that investigated the impact of kneading on the porosity of alumina support through paste composition or operating conditions, have focused on extruded and calcinated alumina supports \cite{Walendziewski:1994,Landers_2010,Zheng_2014,Danner:1987,Dubois:1995}. Therefore, the sole impact of the kneading step preceding extrusion and calcination has not been addressed.

Here, we aim to focus solely on the kneading process to comprehensively understand the impact of both mechanical and physico-chemical aspects on the textural properties of boehmite pastes. We first demonstrate that, for a fixed chemical composition, the temporal evolution of the torque recorded throughout the kneading process, as well as the paste textural properties, are controlled by the deformation accumulated during kneading. In addition, we show the key role of the paste pH through the addition of acid and basic aqueous solutions with various concentrations. We further introduce the product of the accumulated deformation, $\gamma$, and the neutralization ratio, $t_b$, defined as the molar ratio between base and acid added to the boehmite paste, as an empirical control parameter accounting for both mechanical effects and physico-chemical processes during kneading. We show that the parameter $\gamma\times t_b$ allows one to capture remarkably well the evolution of textural properties of boehmite pastes produced by kneading.

The outline of the paper is as follows. In Sect.~\ref{sec:MM}, we introduce the method followed for preparing boehmite pastes, and the pilot kneader used for that purpose, as well as the analytical techniques applied for characterizing the textural properties of the pastes. Section~\ref{sec:Results} reports our main findings and discusses how the kneading operating conditions and the composition of the paste affect its porosity and other parameters measured at various stages of the kneading process. Finally, we discuss our results and conclude in Sect.~\ref{sec:Discussion and conclusion}.

\section{Materials and methods}
\label{sec:MM}

\subsection{Boehmite paste preparation}
\label{sec:PastePrep}
The present experimental study focuses on boehmite pastes prepared using Pural SB3 (Sasol) powder. This boehmite powder consists of micrometric grains, whose size ranges from 1 to 100~$\mu$m, as determined by scanning electron microscopy \cite{Sudreau_2023}. These grains, referred to as \textit{agglomerates} in the literature, are assemblies of \textit{aggregates} sized around hundreds of nanometers. These aggregates themselves are composed of stacked nanometric particles termed \textit{crystallites}, typically measuring between 10 and 20~nm in length and 3 to 10~nm in width and thickness \cite{Sudreau_2023,Gallois_2016,Morin_2014}. This boehmite powder has a specific surface area of $260~\rm m^2.g^{-1}$, as computed with the Brunauer, Emmett, and Teller (BET) model, and a mesoporous volume of $0.38~\rm mL.g^{-1}$, as determined by nitrogen sorption. Additionally, mercury intrusion porosimetry reveals a macroporous volume of about $0.19~\rm mL.g^{-1}$ (see Fig.~\ref{fig:Fig_9} in Appendix~\ref{app:Annexe_Poudre} for details on the pore size distributions). In practice, the boehmite powder is stored at ambient temperature and contains about 14$\%$wt.~of humidity, as determined indirectly during the dehydration process of boehmite into alumina.

\begin{figure}[t!]
    \centering
    \includegraphics[width=1\linewidth]{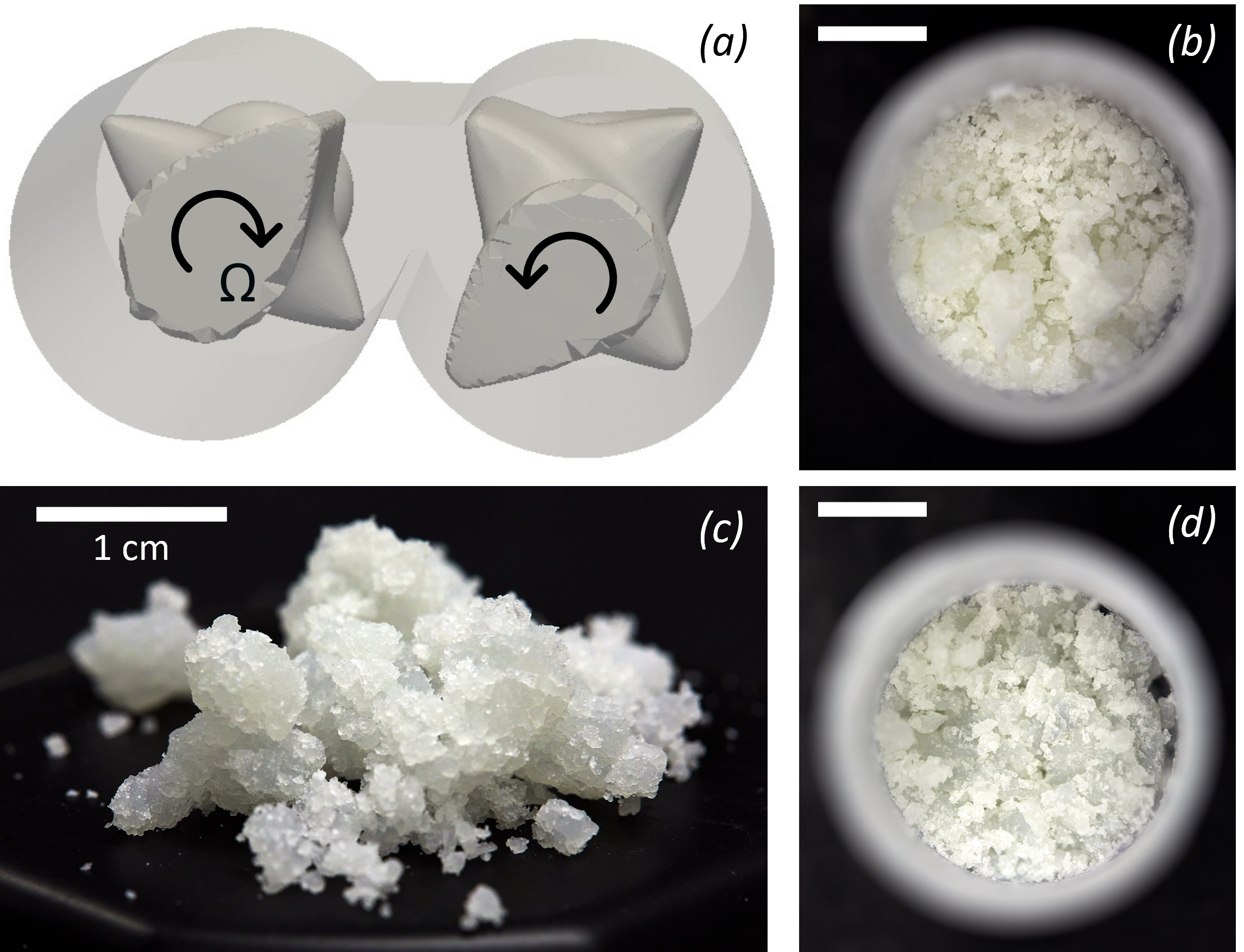}
    \caption{\label{fig:Fig_1bis} (a) Sketch of the laboratory pilot kneader. The dimensions of the kneading cell are 40 x 80~mm, with a usable volume of $80~\rm cm^3$. Pictures of the granular boehmite pastes: (b) peptized and (c),(d) neutralized samples. White bars set the scale in each picture and correspond to 1~cm.}
\end{figure}

Boehmite pastes are prepared into an 80~cm$^3$ laboratory pilot kneader (Pastograph, Brabender) equipped with two contrarotating cam blades with a velocity ratio of $2/3$ [see sketch in Fig.~\ref{fig:Fig_1bis}(a)]. In the present work, velocities refer to the fastest blade, although in practice, the slowest blade is the one directly connected to the motor. The pilot device is also equipped with a force sensor (HBM) for torque measurement and a Pt100 sensor to monitor the temperature inside the kneading cell.

The standard kneading protocol begins with introducing 30.23~mL of a 3.7$\%$wt.~nitric acid solution (Thermo Scientific) into the kneader, initially at rest. Subsequently, the kneader is turned on at $\Omega_A=50$~rpm, and the boehmite powder is progressively introduced in four batches of 9.5~g each, added every minute. This step of total duration $\Delta t_P=4~\rm min$ and referred to as the \textit{peptization} phase, allows for the dispersion of powder agglomerates by inducing electrostatic repulsion through proton adsorption on the surface of the boehmite crystallites \cite{Karouia_13,Speyer_2020,Zheng_2014,Fauchadour_2002}. The result is the formation of an acidic granular paste, shown in Fig.~\ref{fig:Fig_1bis}(b), whose pH is around $4.3$ (as measured with a specific semi-solid probe from HANNA Instruments). This acidic paste, hereafter referred to as peptized paste, is kneaded at $\Omega_A=50~\rm rpm$ during $\Delta t_A=30~\rm min$. 
Following the ``acid kneading'' phase, 11.74~mL of a 1.0$\%$wt.~ammonia solution (LCH Chimie) is introduced into the kneader with a syringe pump at a controlled flow rate of $4.11~\rm mL.min^{-1}$. The injection flow rate corresponds to a duration $\Delta t_N \simeq 3~\rm min$, which is the time practically required to achieve a homogeneous and pasty texture. Note that using faster injection rates leads to a stickier paste that adheres to the blades and cannot be thoroughly mixed in the kneader, even before the entire amount of liquid has been incorporated. The concentration of the ammonia solution is fixed by the neutralization ratio $t_b$, which is defined as the molar ratio between base and acid added to the boehmite powder. In the standard protocol, the neutralization ratio is set at $40\%$, which results in a paste with a pH = 6.2. This phase of ammonia addition is referred to as the \textit{neutralization} phase. It leads to the screening of the surface charges of boehmite crystallites, consequently enhancing the cohesion of the paste, while increasing its porosity \cite{Karouia_13}. Finally, the resulting boehmite paste undergoes a last kneading phase, called ``basic kneading'' and performed at $\Omega_B =50$~rpm during $\Delta t_B = 2$~h. The final neutralized paste, illustrated in Figs.~\ref{fig:Fig_1bis}(c) and \ref{fig:Fig_1bis}(d), has a fixed mass fraction of 41$\%$wt.~of boehmite.

The blade rotational speed, the kneading duration, and the neutralization ratio used for the standard protocol described above are varied to assess their impact on the properties of the boehmite paste. In practice, the various mixing speeds remain constant during each phase, varying between 10 and 100~rpm, but the mixing speeds $\Omega_A$ and $\Omega_B$, during the acid and basic kneading phases, respectively, may differ. Moreover, the duration of the acid kneading phase ranges from $\Delta t_A=3$~min to 25~min, while that of the basic kneading phase extends from $\Delta t_B=3$~min to 4~h. Concerning the paste composition, the neutralization rate is varied from $t_b=10\%$ to 130$\%$, through the addition of ammonia solutions whose concentrations range from 0.3 to 3.4$\%$wt. Such variations in $t_b$ result in a paste pH varying from 5 to 9.

\subsection{Textural characterization}
\label{sec:Charac}

In order to characterize the impact of kneading on the properties of boehmite pastes, the textural characteristics of dried boehmite pastes are determined by nitrogen sorption analysis and mercury intrusion porosimetry, two complementary techniques commonly used for characterizing porous materials  \cite{Schlumberger_2021, IUPAC_1994}. Nitrogen sorption allows for the examination of the micro- and meso-porosity, encompassing pores smaller than 50~nm, while mercury porosimetry probes pores with a diameter ranging from 3.7~nm to 400~µm. 
Following the kneading protocol and prior to any porosity measurements, the samples are dried for 20~h at 80$^\circ$C in an oven. Just before textural measurements, samples are heated up to110$^\circ$C for 6~h, under vacuum and at atmospheric pressure for nitrogen and mercury analyses, respectively. The drying step is crucial to remove free water that could obstruct certain pores and potentially alter the measurements \cite{Bogner:2020}. Note that the drying temperature is kept below 130$^\circ$C to prevent the conversion of boehmite into alumina \cite{Trimm_1986}. We checked that the porosimetry results are not sensitive to the drying conditions by analyzing samples from the same paste after drying at 40$^\circ$C, 80$^\circ$C, and 120$^\circ$C for durations of 6, 20, and 24~h, respectively.

\textit{Nitrogen sorption.--} Nitrogen adsorption-desorption isotherms are measured at 77 K with an ASAP 2420 device (Micrometrics) \cite{IUPAC_1994}. The BET model is applied to compute the specific surface area from the nitrogen isotherms \cite{BET_1938}, while the Barrett, Joyner, and Halenda (BJH)~model is used to determine the pore size distribution \cite{BJH_1951}. The mesoporosity assumption within the BJH model is consistent with the observed isotherm shape in this study, in accordance with the IUPAC classification \cite{IUPAC_1985}. The volume of mesopores, encompassing pores with diameters ranging from 2 to 50~nm \cite{IUPAC_2015} as probed by nitrogen, is considered equivalent to the injected nitrogen volume at saturation. This assumption is valid due to the negligible microporous volume of the boehmite samples, as discussed in Sect.~\ref{sec:TotalDeformation}.

\textit{Mercury intrusion porosimetry.--} Mercury intrusion measurements are performed on an Autopore 9500 porometer (Micrometrics) with a pressure range spanning from 3.5~kPa to 413.4~MPa. The mercury volume increments are converted into pore diameters using the conventional Washburn equation computed by considering a surface tension of 0.485~N.m$^{-1}$, and a contact angle of 140$^\circ$ \cite{Washburn_1921}. These calculations allowed us to determine a truncated mesoporous volume --considering only pores with diameters ranging from 3.7~nm and 50~nm due to the important size of mercury atoms-- and a macroporous volume \cite{Giesche_2006}. The uncertainty on nitrogen and mercury analyses is taken as twice the standard deviation of each textural property characterized on $2\times 7=14$ samples from two different pastes of identical composition and prepared following the exact same kneading protocol.

\begin{figure*}[t!]
    \centering
    \includegraphics[width=1\linewidth]{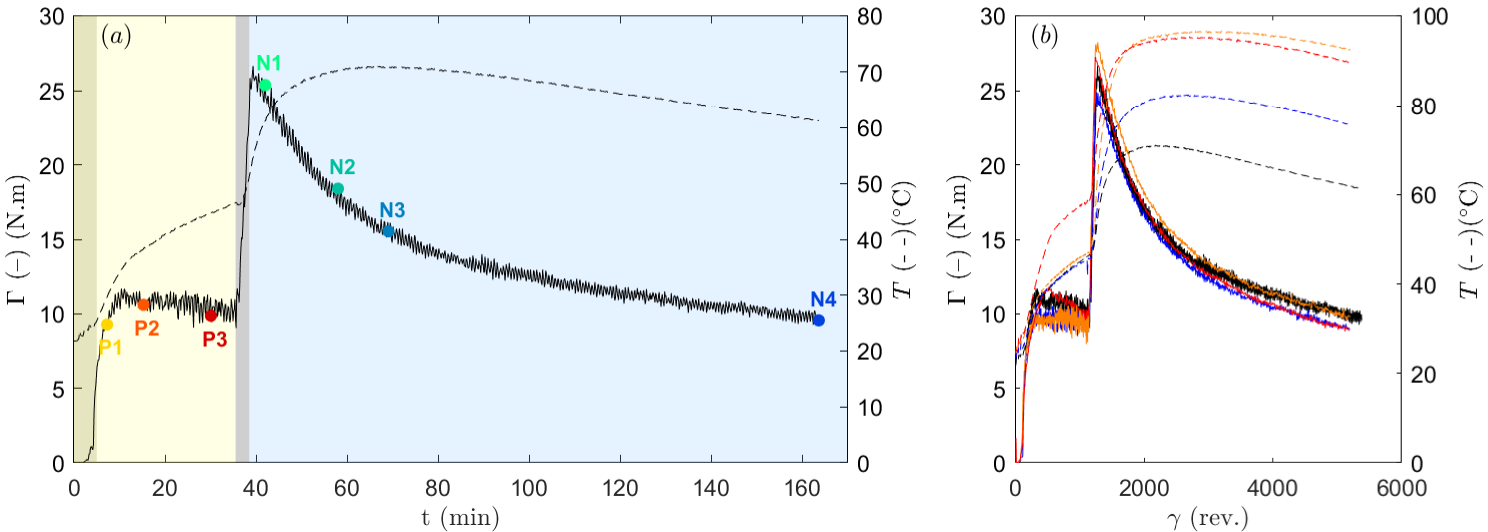}
    \caption{\label{fig:Fig_1} (a) Temporal evolution of the torque $\Gamma$ (continuous line) and temperature $T$ (dashed line) as measured in-situ during a standard kneading experiment performed on a paste neutralized at $t_b=40\%$. The first step of duration $\Delta t_P=4~\rm min$. highlighted in dark yellow corresponds to the peptization, i.e., the formation of a granular paste by mixing the boehmite powder at $\Omega_A=50$~rpm with the acid solution. The peptization is followed by the ``acid kneading'' phase (yellow interval) performed at $\Omega_A=50$~rpm during $\Delta t_A=30$~min. The next time interval $\Delta t_N \simeq 3~\rm min$, highlighted in gray, corresponds to the neutralization phase, i.e., the addition of a basic solution while the mixing speed is kept at $\Omega_A$. The final ``basic kneading'' phase, highlighted in blue, is performed at $\Omega_B=50$~rpm during $\Delta t_B=2$~h. The colored points along the torque curve $\Gamma(t)$ mark the times at which peptized (P1 to P3) and neutralized (N1 to N4) pastes are extracted from the pilot kneader for textural characterization. Note that an additional sample N5 may be extracted at $t=278$~min (see Fig.~\ref{fig:Fig_11} in Appendix~\ref{app:Annexe_Torque}). (b)~Torque $\Gamma$ and temperature $T$ evolution vs.~the deformation $\gamma$ accumulated during the kneading of pastes neutralized at $t_b=40\%$. The colored lines refer to various pairs of rotation speeds ($\Omega_A$, $\Omega_B$) used for the acid and basic kneading phases, respectively: (50~rpm, 50~rpm) (black), (50 rpm, 75 rpm) (blue), (50 rpm, 100~rpm) (orange), and (100~rpm, 100~rpm) (red).}
\end{figure*}

\section{Results}
\label{sec:Results}
The primary goal of the present experimental investigation is to assess how the textural properties of boehmite pastes in both their peptized and neutralized states are affected by kneading speed, duration, and paste composition. We first report on the influence of the kneading speed and duration on pastes neutralized at $t_b=40\%$ (pH = 6.2), where the mixing speed ranges from 50 to 100~rpm, while the total duration varies from 80~min to 158~min. Subsequently, the influence of the composition is quantified on samples with various degrees of neutralization $t_b$, with pH ranging from 5.2 to 8.6. 

\subsection{Influence of the accumulated deformation}
\label{sec:TotalDeformation}

Figure~\ref{fig:Fig_1}(a) depicts the temporal evolution of the torque $\Gamma(t)$ and the temperature $T(t)$ as recorded during the standard kneading process outlined in Sect.~\ref{sec:PastePrep}. The acid kneading is highlighted in yellow, while the basic kneading is indicated in blue. Upon the addition of boehmite powder to the nitric acid solution, which sets the origin of time ($t=0$), the torque $\Gamma$ experiences a rapid increase within the first 5~min as the paste begins to form. Subsequently, $\Gamma$ reaches a plateau value associated with the kneading of a stable granular paste [see Fig.~\ref{fig:Fig_1bis}(b)]. Upon neutralization from $t=35~\rm min$ to $t=38~\rm min$ [see vertical gray stripe in Fig.~\ref{fig:Fig_1}(a)], the torque displays a sharp increase due to the agglomeration of boehmite aggregates, followed by an exponential decay associated with paste homogenization. Throughout the entire kneading process, the temperature $T$ exhibits a significant increase. However, the exact cause behind this temperature increase remains unclear and could be attributed to multiple factors such as viscous dissipation resulting from mixing, acid dilution, and boehmite surface protonation  \cite{Paul_2003,Cazacliu:2008,Tombacz:2001,Sun:2003,Seem:2015}.

The temporal evolution of $\Gamma$ and $T$ described above is significantly impacted by both the mixing speed and duration. Figure~\ref{fig:Fig_1}(b) displays the torque and temperature evolution from four distinct tests performed at various speeds and durations, plotted as a function of the number $\gamma$ of revolutions of the slowest blade, which represents a measure of the total deformation accumulated since the start of the kneading process. Due to the complex geometry of the kneader, a precise estimate of the deformation undergone by the paste is out of reach. Therefore, in the following, we shall simply assimilate $\gamma$ to the accumulated deformation.
Remarkably, all four torque measurements collapse onto a master curve as a function of $\gamma$ (see Fig.~\ref{fig:Fig_10} in Appendix~\ref{app:Annexe_Torque} for the same data as a function of time). This shows that, for a given chemical composition of the paste, $\gamma$
is the relevant control parameter that sets the torque evolution during both the peptization and neutralization phases. As for the temperature, the relevant operating parameter appears to be the blade rotation speed. Indeed, the temperature increases roughly quadratically with the mixing speed for a fixed total deformation. This observation is compatible with the kinetic energy of the rotating blades setting the sample temperature. 

\begin{figure}[t!]
    \centering
    \includegraphics[width=1\linewidth]{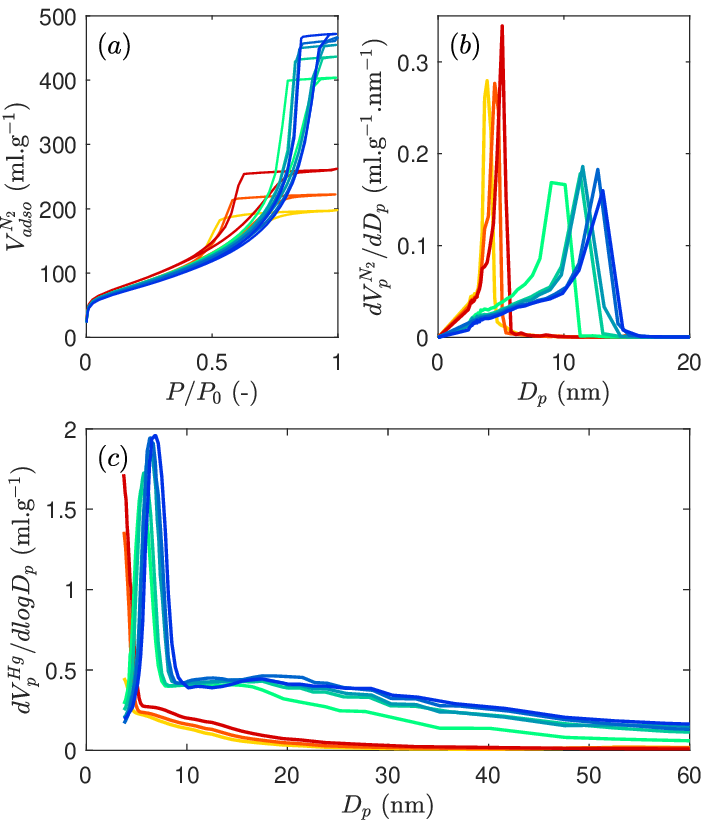}
    \caption{\label{fig:Fig_2} Influence of the accumulated deformation on the pore size distribution of a boehmite paste before and after neutralization at $t_b=40\%$, and prepared at $\Omega_A = \Omega_B =$~50~rpm. (a)~Nitrogen adsorption-desorption isotherms and (b)~pore size distribution computed from the BJH model applied on nitrogen desorption isotherm. (c)~Pore size distribution estimated from mercury porosimetry measurements. Color codes for an increasing level of deformation accumulated by the boehmite paste along the kneading protocol. Samples are extracted at times marked in Fig.~\ref{fig:Fig_1}(a) and labeled P1 to P3 (yellow to red) in the peptization phase and N1 to N5 (cyan to dark blue) in the neutralization phase.}
    \end{figure}

To further quantify the influence of the kneading speed and duration on the textural attributes of boehmite pastes, we extract samples during both the acid and basic kneading phases [see points P1 to P3 and N1 to N4, respectively, in Fig.~\ref{fig:Fig_1}(a)]. The samples are dried before performing nitrogen adsorption and mercury intrusion porosimetry tests following the protocol described in Sect.~\ref{sec:Charac}. The results reported in Fig.~\ref{fig:Fig_2} focus on the impact of the kneading duration on peptized and neutralized pastes prepared at 50~rpm.
The nitrogen adsorption-desorption isotherms for both data sets are shown in Fig.~\ref{fig:Fig_2}(a). It is observed that each isotherm is a type IV with an H2 hysteresis, following the IUPAC classification \cite{IUPAC_1985}. Our observations suggest that all the boehmite samples extracted at various stages of the peptization-neutralization process primarily exhibit mesoporous characteristics, featuring interconnected and/or bottle-shaped pores \cite{IUPAC_1985}. Moreover, based on the results in Fig.~\ref{fig:Fig_2}(a), we can assume that our samples are mainly mesoporous and hold negligible microporosity (pores smaller than 2~nm). In fact, since the nitrogen analysis conducted here covers only pores with diameters exceeding 1.7~nm (and not the entire range of micropores from 0 to 2~nm), the porous distributions tend towards zero at low diameter, suggesting negligible micropore presence. Note that the smallest measurable pore diameter depends on the relative pressure increase applied at each step for isotherm measurement. We can, therefore, safely conclude that isotherms allow us to determine the sample mesoporous volume based on the nitrogen volume adsorbed at saturation ($P/P_0 = 1$), as microporosity is negligible. 
In this context, we can further exploit Fig.~\ref{fig:Fig_2}(a). First, we observe an increase in the porous volume for increasing kneading duration in both the peptization and neutralization phases. Second, the mesoporous volume displays a twofold increase upon the neutralization step. Namely, the mean mesoporous volume of the boehmite paste increases from $226~\rm mL.g^{-1}$ in the peptized state to $445~\rm mL.g^{-1}$ in the neutralized state. 

From nitrogen desorption isotherm and mercury porosimetry analysis, we extract the pore size distribution plotted respectively in Figs.~\ref{fig:Fig_2}(b) and \ref{fig:Fig_2}(c). The nitrogen desorption analysis exploited with the BJH model reveals that the neutralized pastes have larger pores compared to the peptized pastes, a result consistent with the volume increase discussed above. Moreover, both mercury and nitrogen pore size distributions show that longer kneading durations lead to larger pore diameters for both peptized and neutralized pastes. In order to quantify the effect of the kneading duration on the pore diameter, two mean pore diameters are extracted from the pore size distributions, considering their bimodal nature (see Fig.~\ref{fig:Fig_12} in Appendix~\ref{app:Annexe_Deconvolution}).

\begin{figure*}[t!]
    \centering
    \includegraphics[width=1\linewidth]{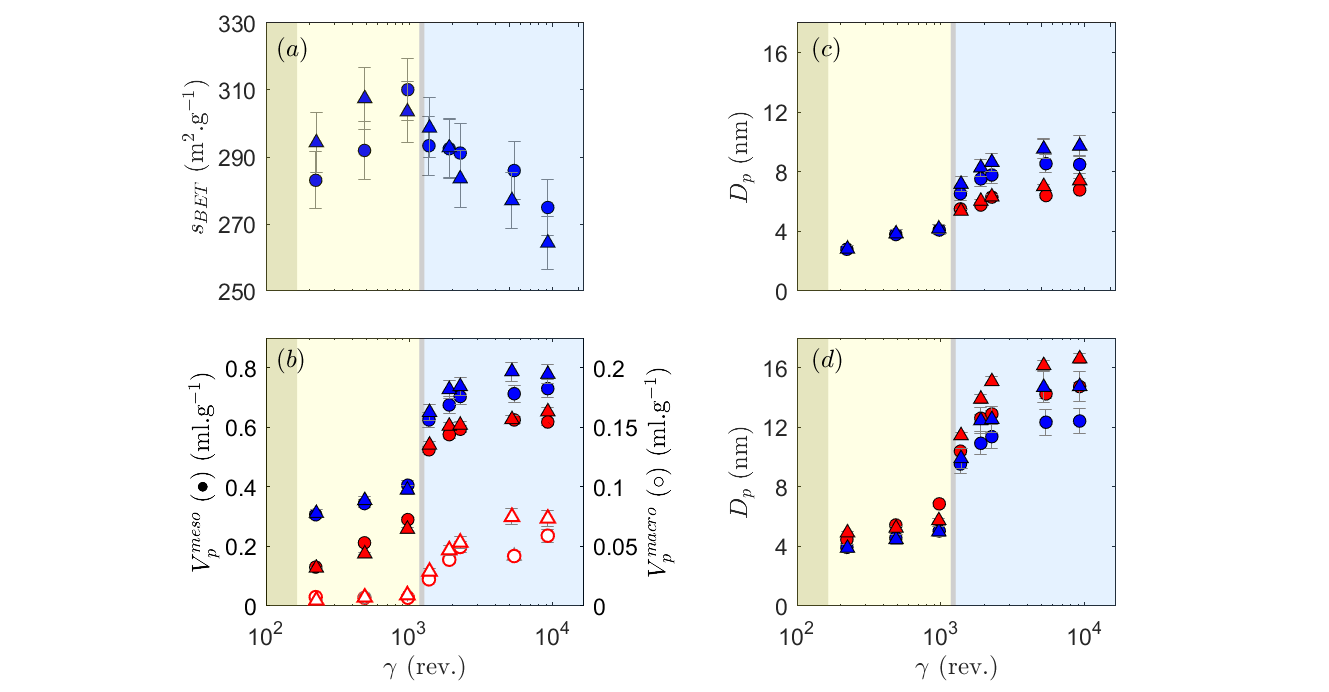}
    \caption{\label{fig:Fig_3} Influence of the accumulated deformation on the textural properties of a paste neutralized at $t_b=40\%$. (a) Specific surface area s$_{BET}$, (b) mesoporous $V^{meso}_{p}$ (full symbols) and macroporous volume $V^{macro}_{p}$ (empty symbols), and mean pore diameter $D_{p}$ of first (c) and second (d) deconvoluted pore populations as a function of the accumulated deformation $\gamma$ (see Appendix~\ref{app:Annexe_Deconvolution} for details regarding the deconvolution). Symbols stand for the blade rotation speed of 50 rpm ($\circ$) and 100 rpm ($\triangle$). The colors of the symbols code for the characterization technique: nitrogen adsorption (blue) and mercury porosimetry (red). Missing points in mercury data result from the technique failing to detect pores with a diameter smaller than 3.7~nm. Error bars indicate the uncertainty as explained in Sect.~\ref{sec:MM}. Colored areas correspond to the various phases of the kneading process as defined in Fig.~\ref{fig:Fig_1}.}
\end{figure*}

\begin{figure*}[t!]
    \includegraphics[width=1\linewidth]{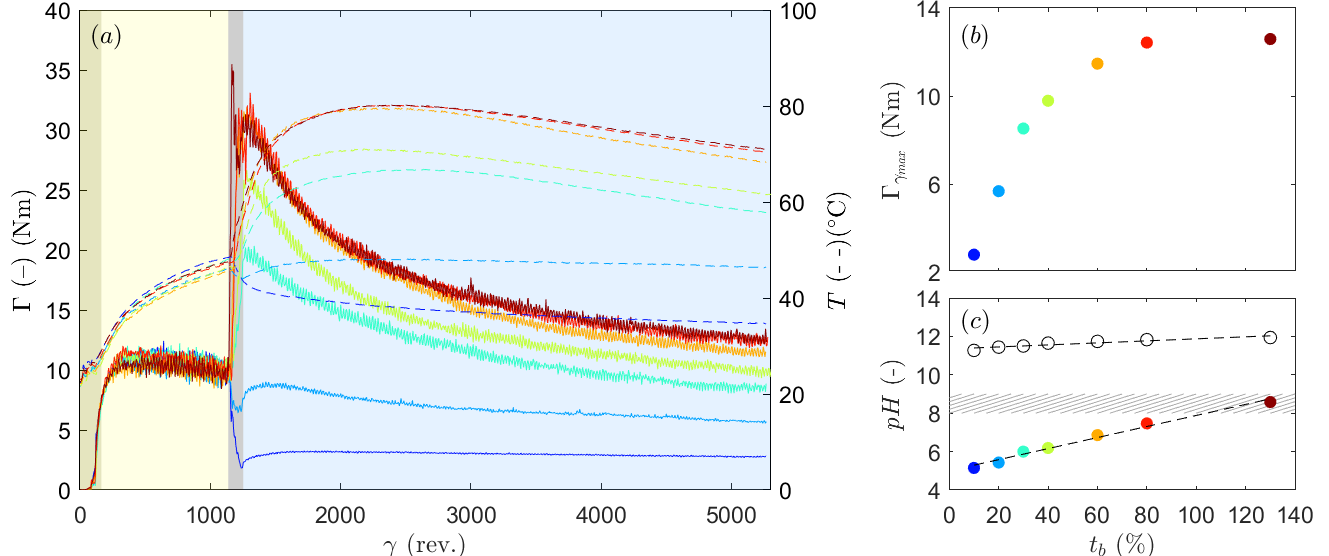}
    \caption{\label{fig:Fig_4} (a) Torque  $\Gamma$ (full line) and temperature $T$ (dashed line) vs.~accumulated deformation for various neutralization ratios $t_b$. (b)~Terminal value of the torque ${\Gamma}_{{\gamma}_{max}}$ reached after the standard kneading protocol presented in Sect.~\ref{sec:MM}, i.e., after 160~min of kneading at $\Omega_A=\Omega_B=50$~rpm, and (c) paste pH (colored points) as a function of the neutralization ratio $t_b$. The point of zero charge is shown as a horizontal striped area (value extracted from Ref.~\cite{Ramsay:1978}). Empty symbols ($\circ$) show the pH of the ammonia solutions used to neutralize the paste for the various values of $t_b$ tested. In (a), (b), and (c), colors code for the pH of the boehmite paste as reported in (c).}
\end{figure*}

Figure~\ref{fig:Fig_3} summarizes the influence of kneading speed and duration on the textural properties of boehmite pastes produced using the different kneading speeds $\Omega_A=\Omega_B=50$ or $\Omega_A=\Omega_B=100~\rm rpm$. Figures~\ref{fig:Fig_3}(a) and \ref{fig:Fig_3}(b) show that the accumulated deformation $\gamma$ is the control parameter for both specific surface area and porous volume evolution, as the data sets obtained at two different speeds fall on to a master curve, up to experimental uncertainty. Additionally, Figs.~\ref{fig:Fig_3}(c) and \ref{fig:Fig_3}(d) show that the total deformation is also the relevant parameter governing the evolution of the mean pore diameter for both pore populations. More generally, the evolution of the mean diameters, mirroring the trend of the porous volume, reveals that the larger pore population is the most impacted by the deformation accumulated over the whole kneading protocol.

Upon comparing the mesoporous volume or the mean pore diameters derived from nitrogen sorption and mercury intrusion measurements, discrepancies in the obtained porosity values at a fixed cumulative deformation become evident, as illustrated in Figs.~\ref{fig:Fig_3}(b--d). Still, both techniques exhibit a consistent trend, and the dissimilarities in absolute value can be attributed to the different ranges of pore diameters accessible with each technique (as explained in Sect.~\ref{sec:Charac}) and to differences in the models used for data analysis \cite{IUPAC_1994}. 
Hence, Fig.~\ref{fig:Fig_3} allows us to conclude that dried boehmite pastes are mainly mesoporous, as throughout the entire kneading process, the mean pore diameters of both pore populations remain within the mesoporous diameter range. Furthermore, the porous volume reported in Fig.~\ref{fig:Fig_3}(b) is dominated by the mesoporous component, with the contribution of macroporous volume (resulting from the tail of the larger pore distribution) remaining below $10\%$ of the total volume. It is also worth noting that, except for its specific surface area, the texture of the paste remains stable after 2~h of basic kneading, whereas the torque still decreases significantly beyond that time, i.e., between neutralized samples labeled N4 and N5 points (see Fig.~\ref{fig:Fig_11} in Appendix~\ref{app:Annexe_Torque}). This observation suggests that the torque decay cannot be explained solely by structural modifications at the porosity scale and that another process is at stake. In the context of cement mixing, similar exponential torque decays have been reported for fresh cement pastes and associated with their homogenization by destruction and dispersion of centimetric agglomerates composed of cement, sand, and gravel \cite{Cazacliu:2008,Cazacliu:2009,MorenoJuez:2017}. Along the same lines, we may assume that the torque decay measured during the kneading of boehmite pastes results from the rearrangement of the paste microstructure at a centimetric scale. To confirm this assumption, mechanical characterization and microscopic imaging should be performed on neutralized samples. Such measurements will allow one to quantify the effect of kneading on paste structures at scales exceeding that of the porosity.

\begin{figure*}[t!]
    \centering
    \includegraphics[width=1\linewidth]{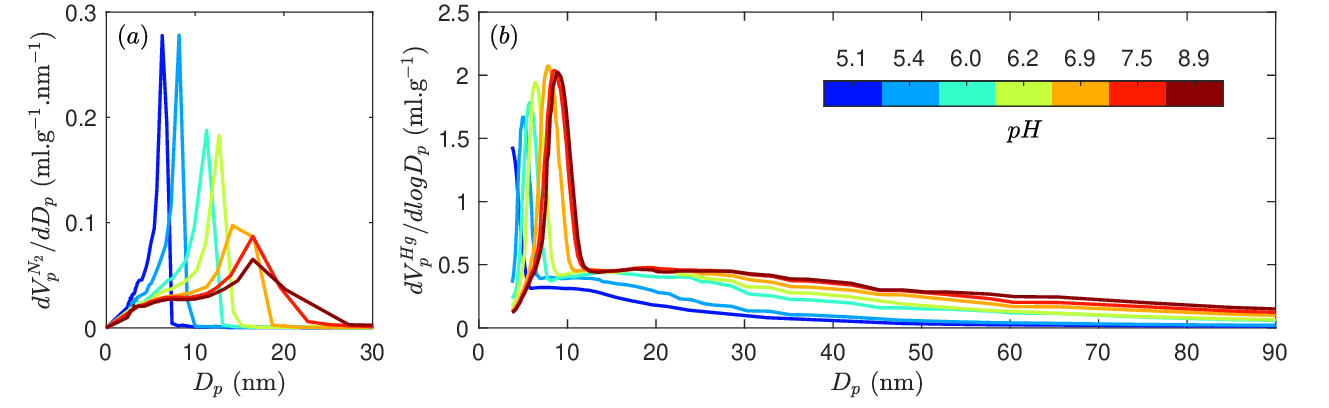}
    \caption{\label{fig:Fig_5} Influence of the pH of boehmite pastes on the pore size distributions obtained from (a) nitrogen adsorption measurements exploited with BJH model and (b) mercury porosimetry analysis. The pastes were prepared at different neutralization ratios $t_b$ following the standard protocol [{see Sect.~\ref{sec:MM}}] and sampled after a deformation of $\gamma=5260~\rm rev.$ [equivalent to the point labeled N4 on Fig.~\ref{fig:Fig_1}(a)].}
\end{figure*}

\begin{figure*}[t!]
    \centering
    \includegraphics[width=1\linewidth]{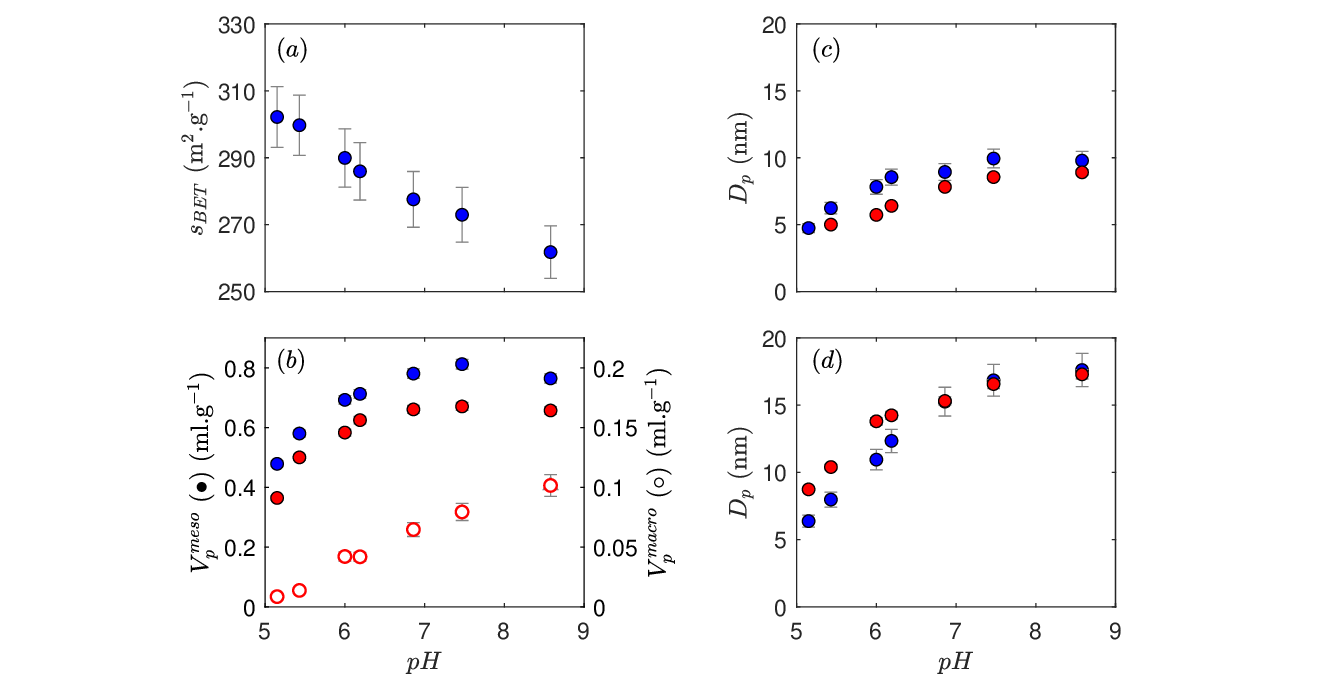}
    \caption{\label{fig:Fig_6} Influence of the paste pH on the textural properties of dried samples. (a) Specific surface area s$_{BET}$, (b) mesoporous $V^{meso}_{p}$ (full symbols) and macroporous volume $V^{macro}_{p}$ (empty symbols), and mean pore diameter $D_{p}$ of (c) first and (d) second deconvoluted pores populations a function of the paste pH. Colors code for the characterization technique: nitrogen adsorption (blue) and mercury porosimetry (red).}
\end{figure*}

\subsection{Influence of the composition of the boehmite paste}
\label{sec:Compo}

We now investigate the effect of the paste composition, more specifically, the effect of the paste pH on both the torque evolution and the structural properties of the paste. Experiments are conducted on neutralized pastes prepared using the standard protocol described in Sect.~\ref{sec:Charac}, where the total kneading duration is 160~min at $\Omega_A=\Omega_B=50$~rpm. The paste composition is modified in the neutralization phase by adding 11.74~g of basic ammonia solution, whose pH varies from 11.3 to 11.9 [see empty symbols in Fig.~\ref{fig:Fig_4}(c)], to make a paste of pH varying from 5.2 to 8.7. Here, it is worth noting that the boehmite mass fraction of all the pastes is constant as a fixed mass of ammonia solution is added. We emphasize that small variations of the pH of the ammonia solution from 11.3 to 11.9 induce large changes in the paste pH from 5.2 to 8.7. 
In practice, we observe that the paste composition has two effects on the torque and temperature measured in-situ, as reported in Fig.~\ref{fig:Fig_4}(a). First, the neutralization ratio $t_b$, or equivalently the paste pH [see filled symbols in Fig.~\ref{fig:Fig_4}(c)], strongly affects the torque evolution, which displays two distinct behaviors depending on the pH of the basic solution. Indeed, for $t_b<30\%$, i.e., upon injecting an ammonia solution with a pH smaller than 11.5, yielding a paste of final pH smaller than 6.0, induces a sudden drop in torque. In contrast, for $t_b\ge 30\%$n, i.e., upon adding a basic solution with a pH larger than 11.5, triggers a rapid increase in $\Gamma(t)$. This dual behavior arises from the competition between the paste dilution resulting from the addition of the basic solution and the aggregation/agglomeration of the boehmite aggregates, promoted by the pH increase. 
Such a competition is also reflected by the terminal value of the torque at the end of the basic kneading phase, i.e., after ${\gamma}=5260$ revolutions, which shows a strong increase with the neutralization ratio $t_b$, up to a plateau value of about 12.5~N.m reached for a neutralization ratio $t_b\simeq$~80$\%$ [see Fig.~\ref{fig:Fig_4}(b)]. The torque increase can be explained as a macroscopic response triggered by the enhancement of the aggregation of boehmite crystallites induced at the nanoscale by the pH increase. Interestingly, the saturation of the torque occurs at a pH lower than the isoelectric point, for $t_b\simeq 80\%$ [see striped area in Fig.~\ref{fig:Fig_4}(c)]. This effect could be due to the confinement of the boehmite crystallites that are densely packed within the pastes.

\begin{figure*}[t!]
    \centering
    \includegraphics[width=1\linewidth]{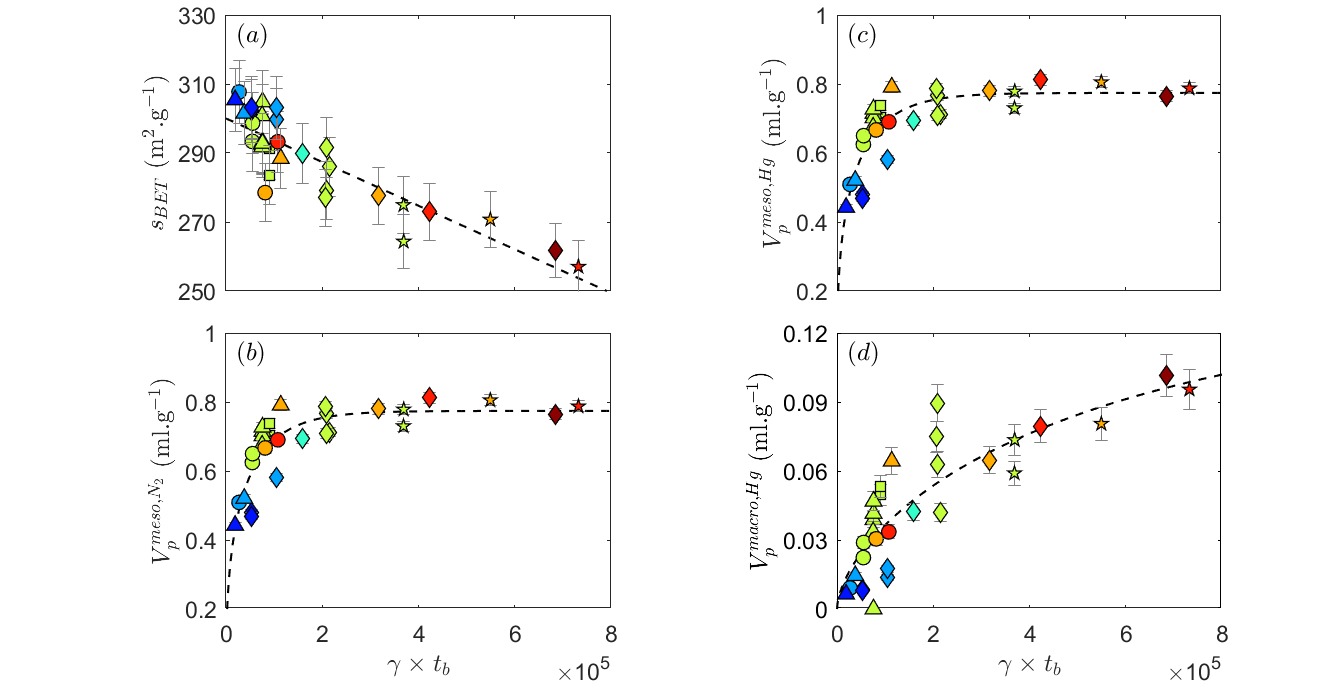}
    \caption{\label{fig:Fig_7}Influence of the product of total deformation accumulated during kneading $\gamma$ and the neutralization ratio $t_b$ on textural properties : (a) specific surface area s$_{BET}$, mesoporous volume analyzed in (b) nitrogen adsorption $V^{meso,N2}_{p}$ and nitrogen porosimetry $V^{meso,Hg}_{p}$ (c) and (d) macroporous volume $V^{macro}_{p}$. Colors code for the neutralization ratio $t_b$ and the symbols shapes code for the total deformation : 1350~rev. ($\circ$), 1900~rev. ($\triangle$), 5260~rev. ($\diamond$), 9160~rev. ($\star$). Dashed curves are guides for the eye : linear in (a) and stretched exponential in (b), (c), and (d).}
\end{figure*}

Next, we examine the impact of paste composition on the textural properties of neutralized pastes. To this aim, we collected pastes prepared at different pH levels and measured the pore size distributions computed from nitrogen sorption and mercury porosimetry. Figure~\ref{fig:Fig_5} shows a widening of the bimodal distribution towards larger mesopores for increasing pH values. More precisely, nitrogen measurements shown in Fig.~\ref{fig:Fig_5}(a) reveal the appearance of larger but less numerous pores as the pH increases \footnote{The number of pores of size $D_p$ can be estimated as the ratio of the porous volume associated with the pore diameter $D_p$ in the pore size distribution to the volume of either a spherical pore of diameter $D_p$ or a cylindrical pore with diameter $D_p$ and typical length 20~nm. We checked that both estimates yield the same trends for the evolution of the number of pores.}. Similarly, mercury porosimetry shows an increase in the size of pores [see Fig.~\ref{fig:Fig_5}(b)]. The widening of both pore size distributions is consistent with the evolution of additional textural properties reported in Fig.~\ref{fig:Fig_6}. Indeed, the specific surface area computed from the nitrogen sorption experiments decreases linearly with the pH [see Fig.~\ref{fig:Fig_6}(a)], in agreement with the shift towards bigger but fewer pores shown in Fig.~\ref{fig:Fig_5}(a). Moreover, both meso- and macroporous volumes presented in Fig.~\ref{fig:Fig_6}(b) increase for increasing pH values and yet follow two different trends. The mesoporous volume increases weakly until pH = 6.9, before saturating around $0.78~\rm mL.g^{-1}$ according to nitrogen analysis (compared to $0.66~\rm mL.g^{-1}$ measured by mercury intrusion porosimetry). In contrast, the macroporous volume shows a linear increase with pH. Therefore, boehmite pastes become progressively more porous with higher neutralization ratios. This increase in pore volume coincides with an increase in the mean pore size diameter for both distributions, as seen in Figs.~\ref{fig:Fig_6}(c) and \ref{fig:Fig_6}(d). We also emphasize that the mean diameter of the large pore population is more significantly affected than that of the smaller pore population, as observed in Sect.~\ref{sec:TotalDeformation}.

\section{Discussion and conclusion}
\label{sec:Discussion and conclusion}

Let us now summarize and discuss the key findings from this study. We have reported on the influence of both kneading operating conditions (speed and duration) and paste composition on in-situ measurements of the torque and temperature in a pilot kneader, and on the paste textural properties measured ex-situ. The mechanical effect of kneading is completely captured by the accumulated deformation, while the paste pH stands out as a significant parameter regarding paste composition. Comparatively, the paste pH has more effect on textural properties than the accumulated deformation. For instance, when analyzing the mesoporous volume (via nitrogen adsorption), neutralized pastes exhibit an increase from 0.48 to $0.78~\rm mL.g^{-1}$ with a rise in pH from 5.2 to 6.9 [see Fig.~\ref{fig:Fig_6}(b)]. However, this volume only increases from $0.62$ to $0.71~\rm mL.g^{-1}$ for a two-hour increase in the duration of basic kneading at pH = 6.2 [see the difference between the first and fourth points in Fig.~\ref{fig:Fig_3}(b)].

Furthermore, a deeper analysis allows us to rationalize both the mechanical and the physico-chemical effects on textural properties using a single parameter. Indeed, Fig.~\ref{fig:Fig_7} presents the specific surface area s$_{BET}$, mesoporous volume $V^{meso}_{p}$ and macroporous volume $V^{macro}_{p}$, and mean pore diameters $D_{p}$ as a function of the total deformation $\gamma$ multiplied by the neutralization ratio $t_b$. Remarkably, this empirical control parameter allows collapsing all the data characterizing the textural properties onto master curves  (see Fig.~\ref{fig:Fig_13} in Appendix~\ref{app:Annexe_Descriptors} for textural properties reported as a function of $\gamma$ and $t_b$ separately). Let us emphasize that  Fig.~\ref{fig:Fig_7} contains the textural properties of 32 neutralized pastes. Sixteen of them have already been presented in Sect.~\ref{sec:Results}, i.e., 10 data points come from the deformation evolution data set pictured in Fig.~\ref{fig:Fig_4}, while 7 points correspond to the composition database reported in Fig.~\ref{fig:Fig_6}, and one point at $t_b$~=~40$\%$ is common to both data sets. 
The other 16 pastes were prepared with 16 different combinations of pastes pH (from 5.2 to 8.5), basic kneading time (from 3~min to 4~h), and mixing speed (10, 25, 50, 75, and 100~rpm) to cover the broadest possible range of operating conditions.

Although the parameter $\gamma\times t_b$ remains purely phenomenological, its identification is valuable for industrial applications, especially for making catalyst supports, because it demonstrates the possibility of achieving a broad range of textural properties from the same boehmite powder. Moreover, this parameter highlights the coupling between two independent control parameters --mechanical and physico-chemical-- which can be adjusted to tune textural properties by modifying the kneading operating conditions.
At the laboratory scale, such a control parameter for textural properties holds value in optimizing new paste characteristics. Indeed, the validity of this empirical parameter can be evaluated with just a few tests on new supports, which significantly reduces the number of tests required to identify trends. Subsequently, this parameter can be used as a guide to define optimal operating conditions for achieving a targeted porosity for catalytic tests. 
However,  it remains crucial to explore the applicability of this empirical parameter to other pastes and protocols in future works. Indeed, numerous parameters were kept fixed in the present study, such as the type of boehmite, the solid and nitric acid concentrations, and the shape of kneading blades, which may all also affect the texture of boehmite pastes. 

More generally, it is interesting to revisit our results in the framework of the existing literature. When considering the torque evolution during the kneading step of the extrudate manufacturing process --a step that has received limited attention--, we refer to Ref.~\cite{Danner:1987}, where it is noted that the time to reach a given torque level in a 3-L kneader decreases linearly with the mixing speed. This observation is consistent with our findings, which demonstrate a torque decrease driven by the accumulated deformation. 
Regarding textural properties, as recalled in the introduction, previous works on the effects of kneading on alumina porosity have only characterized dried and calcinated extrudates \cite{Walendziewski:1994,Landers_2010,Zheng_2014,Danner:1987,Dubois:1995}. Moreover, these studies employed boehmite powders and peptizing agents that are different from the raw materials used in the present study. Because both raw material composition and thermal treatments impact absolute values of porosity \cite{Alphonse:2005,Trimm_1986,Pourcel_2007}, our ability to make direct comparisons is limited.  Nonetheless, we can compare trends and check that our results are consistent with the literature \cite{Dubois:1995,Landers_2010}. For instance, an increase in the kneading duration leads to a widening of the bimodal pore size distribution of alumina catalytic supports, consequently resulting in an increase in the mean pore diameter for both pore populations. 
Finally, we may discuss the sequential effect of peptization and neutralization steps on boehmite pastes. Seminal Refs.~\cite{Trimm_1986,Danner:1987} report that the peptization step induces a considerable decrease in the macroporous population, while the mesoporous volume is almost not affected. In excellent agreement, our results show that the initial macroporous volume of the boehmite powder is about $0.19~\rm mL.g^{-1}$ (see Fig.~\ref{fig:Fig_9} in Appendix~\ref{app:Annexe_Poudre} for the textural properties of the dry powder) and drops significantly to $0.004~\rm mL.g^{-1}$ just after the peptization step [see the first empty symbols in the yellow interval on Fig.~\ref{fig:Fig_3}(b)]. In contrast, the mesoporous volume analyzed by nitrogen adsorption only decreases marginally from $0.38~\rm mL.g^{-1}$ for the powder to $0.31~\rm mL.g^{-1}$ for the first peptized sample [refer to the first colored points on Fig.~\ref{fig:Fig_3}(b) in the yellow interval]. 

To conclude, this experimental study allowed us to disentangle the effects of the paste pH from that of the accumulated deformation on the textural properties of boehmite pastes prepared by kneading. An outstanding question remains to determine the influence of the kneading preparation step on the final mechanical properties of boehmite pastes. In practice, future experiments should include rheological measurements for determining the linear viscoelastic properties of the paste and micro-indentation to determine their local mechanical properties and, therefore, provide a spatially resolved picture of their elastic properties along the kneading process. 

\begin{acknowledgments}
The authors acknowledge A.~Dandeu, E.~Lécolier, J.-M.~Schweitzer, and M.~Servel for fruitful discussions, as well as N.~Alonso, G.~Canzian, M.~Collaudin, F.~Georget, J.~Jacquemet, D.~Roux, N.~Talbi, and J.~Vivas Dias from the IFPEN analytical department for their help with texture characterization experiments.
\end{acknowledgments}

\appendix

\section{Powder textural properties}
\label{app:Annexe_Poudre}
\begin{figure}[h!]
    \centering
    \includegraphics[width=1\linewidth]{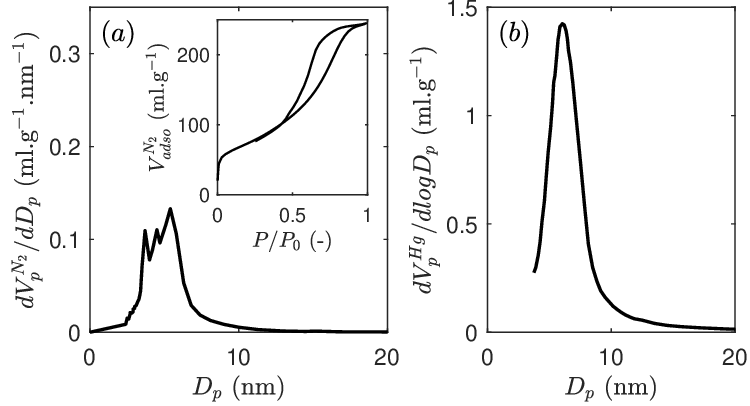}
    \caption{\label{fig:Fig_9} (a) Pore size distribution computed from the BJH model applied on nitrogen desorption isotherm, shown as an inset. (b) Pore size distribution estimated from mercury porosimetry measurements.}
\end{figure}

The boehmite powder textural properties are measured with nitrogen sorption and mercury intrusion porosimetry, according to protocols introduced in Sect.~\ref{sec:MM}. Figure~\ref{fig:Fig_9} presents the nitrogen adsorption-desorption isotherm and the pore size distribution computed from nitrogen and mercury analyses.
Nitrogen sorption analysis measures a mesoporous volume of the dry powder equal to $0.38~\rm mL.g^{-1}$. The mercury porosimetry analysis gives $0.33$ and $0.19~\rm mL.g^{-1}$ for meso- and macro-porous volumes, respectively. 

\begin{figure}[h!]
    \centering
    \includegraphics[width=1\linewidth]{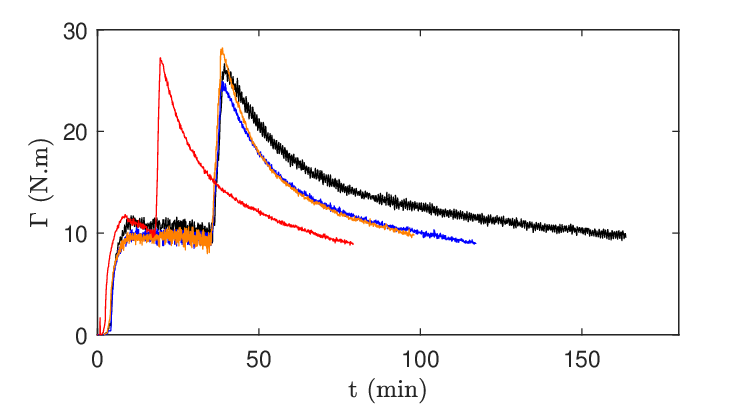}
    \caption{\label{fig:Fig_10} Time-evolution of the torque measured during the kneading of pastes neutralized at 40$\%$. The colors refer to various pairs of rotation speeds ($\Omega_A$, $\Omega_B$) used for the acid and basic kneading phases, respectively: (50~rpm, 50~rpm) (black), (50 rpm, 75 rpm) (blue), (50 rpm, 100~rpm) (orange), and (100~rpm, 100~rpm) (red).}
\end{figure}

\section{Temporal evolution of the torque}
\label{app:Annexe_Torque}

Figure~\ref{fig:Fig_10} shows the same data as in Fig.~\ref{fig:Fig_1}(b) as a function of time, namely the temporal evolution of the torque recorded during four tests performed with various pairs of rotation speeds ($\Omega_A$, $\Omega_B$) used for the acid and basic kneading phases, respectively, and various durations adjusted to reach a constant total deformation of 5260 revolutions.

\begin{figure}[ht]
    \centering
    \includegraphics[width=1\linewidth]{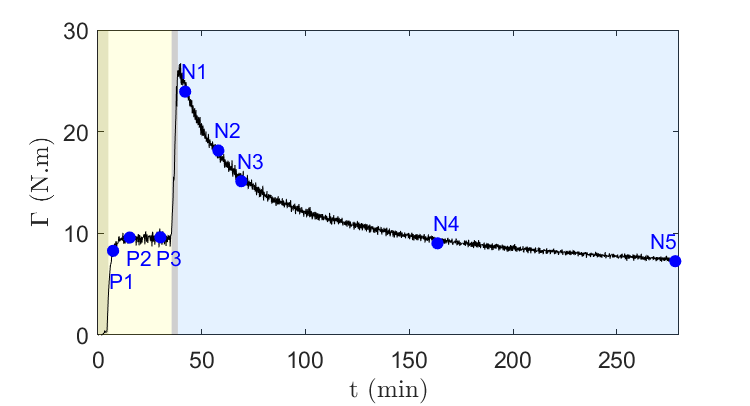}
    \caption{\label{fig:Fig_11}Time-evolution of the torque $\Gamma$ measured during the longest kneading experiment performed with a rotation speed of $\Omega_A=\Omega_B=50~\rm rpm$ to prepare a paste neutralized at $t_b=40\%$. The basic kneading duration was increased to $\Delta t_B=4~\rm h$ while the previous phases were kept as described in the standard protocol (see Sect.~\ref{sec:PastePrep}). The blue points along the torque curve $\Gamma(t)$ indicate the times at which peptized (P1 to P3) and neutralized (N1 to N5) pastes are extracted from the pilot kneader for textural characterization.}
\end{figure}

Figure~\ref{fig:Fig_11} shows the temporal evolution of the torque measured during the longest kneading experiment performed on a paste neutralized at $t_b=40\%$. Indeed, the duration of the basic kneading phase was increased to $\Delta t_B=4$~h when the rotation speed was kept at $\Omega_A=\Omega_B=50$~rpm. This figure emphasizes the continuous decrease of the torque after the end of the standard protocol (see blue point labeled N4 on Fig.~\ref{fig:Fig_11}), as the torque measured after 4~h of basic kneading (labeled N5 on Fig.~\ref{fig:Fig_11}) is significantly lower.

\section{Deconvolution of pore size distributions}

\begin{figure}[h!]
    \centering
    \includegraphics[width=1\linewidth]{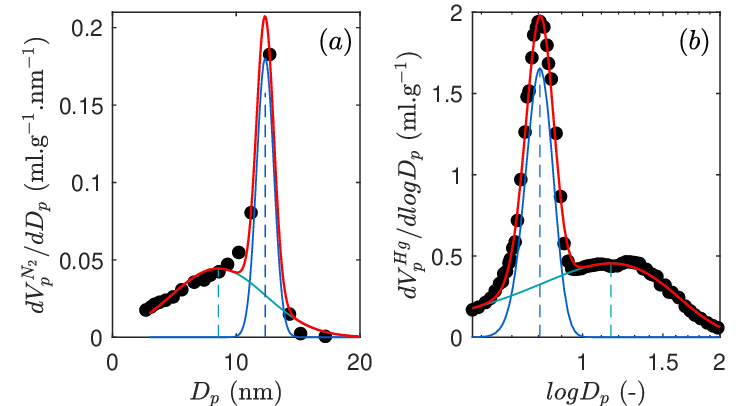}
    \caption{\label{fig:Fig_12} Deconvolution of pore size distribution computed from (a) nitrogen desorption isotherm and (b) mercury porosimetry. Experimental data (black points) are deconvoluted by two Gaussian functions, outlined by the blue curves, whose sum is represented by the red curve. Blue vertical bars indicate the mean value of each Gaussian function, which defines the mean diameter for the small and large pore populations. The distribution computed from nitrogen isotherms is fitted with two Gaussian functions characterized by means of $7.79~\rm nm$ and $11.37~\rm nm$, and standard deviations of $2.6~\rm nm$ and $0.63~\rm nm$, respectively. Similarly, mercury porosimetry distribution is fitted with two Gaussian functions with means of $0.80~\rm nm$ and $1.1~\rm nm$, and standard deviation of $0.08~\rm nm$ and $0.63~\rm nm$, respectively.}
\end{figure}

The paste porosity is characterized by five textural properties: specific surface area, meso- and macroporous volumes, and two mean pore diameters. Those mean pore diameters are extracted from the pore size distribution computed from both nitrogen and mercury porosimetry measurements by fitting a sum of two Gaussian distributions. An example of deconvolution is shown in Fig.~\ref{fig:Fig_12}, for a $t_b=40\%$ neutralized paste kneaded for 160~min at $\Omega_A=\Omega_B=50~\rm rpm$.

\label{app:Annexe_Deconvolution}

\section{Empirical control parameter}
\label{app:Annexe_Descriptors}

\begin{figure*}[ht!]
    \centering
    \includegraphics[width=1\linewidth]{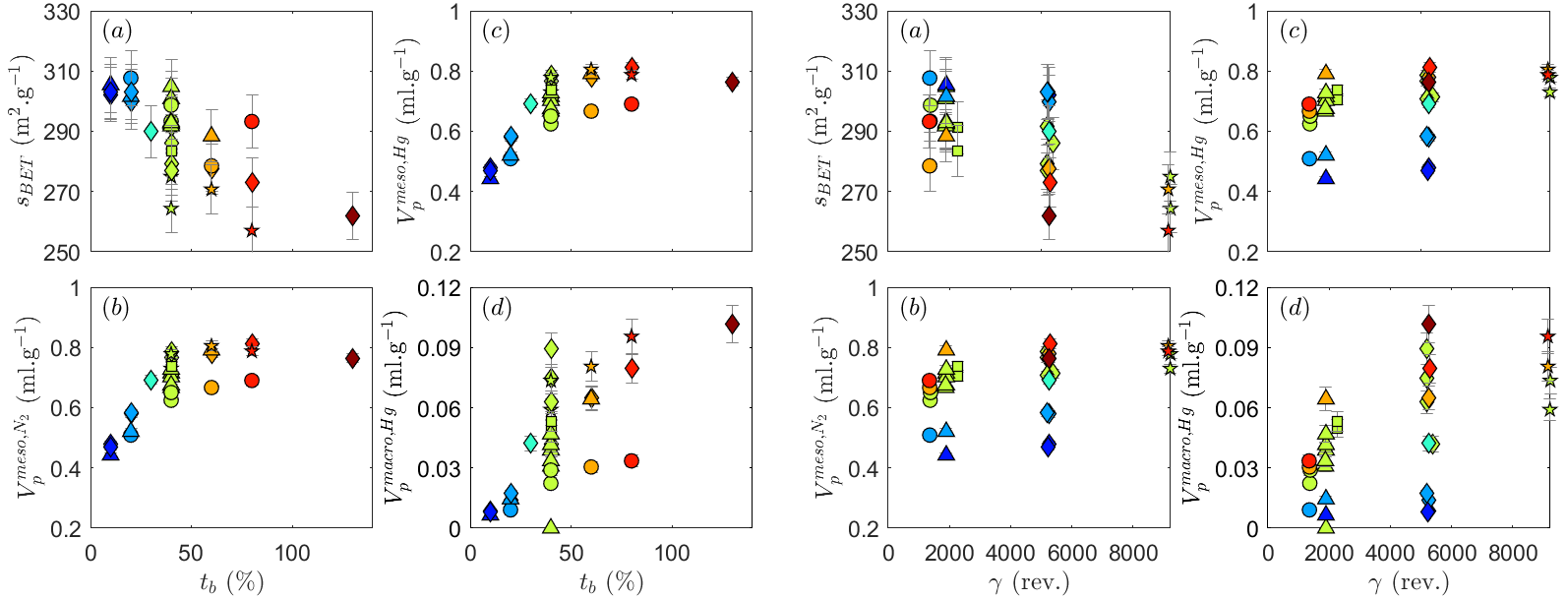}
    \caption{\label{fig:Fig_13} Evolution of textural properties characterized on neutralized pastes as a function of (a)--(d) the neutralization rate $t_b$  and (e)--(f) the accumulated deformation. The product of those two parameters is a relevant control parameter for textural properties, as shown in Fig.~\ref{fig:Fig_7} in the main text. Colors and shapes code the neutralization rate and the total deformation, respectively.}
\end{figure*}

In Sect.~\ref{sec:Discussion and conclusion}, we report that the product of the total accumulated deformation $\gamma$ and the neutralization ratio $t_b$ is a relevant control parameter for textural properties evolution (see Fig.~\ref{fig:Fig_7}). Figure~\ref{fig:Fig_13} presents the evolution of the same textural properties as a function of the total deformation $\gamma$ and the neutralization ratio $t_b$, separately. This figure shows that the product of those parameters is a relevant control parameter as none of them, taken individually, does reveal any clear trend in the evolution of the textural properties.
Moreover, Figs.~\ref{fig:Fig_13}(a)--(d) illustrate that a modification of the paste composition allows one to cover a broader range of textural properties than that obtained by modifying the mechanical operating conditions. Indeed, the texture variations induced by modifying the kneading speed or duration, evaluated at a fixed neutralization ratio, are smaller than those measured by modifying the pH for the three textural properties presented in Fig.~\ref{fig:Fig_13} (specific surface area and meso- and macroporous volumes).

\nocite{*}


\begin{thebibliography}{66}%
\makeatletter
\providecommand \@ifxundefined [1]{%
 \@ifx{#1\undefined}
}%
\providecommand \@ifnum [1]{%
 \ifnum #1\expandafter \@firstoftwo
 \else \expandafter \@secondoftwo
 \fi
}%
\providecommand \@ifx [1]{%
 \ifx #1\expandafter \@firstoftwo
 \else \expandafter \@secondoftwo
 \fi
}%
\providecommand \natexlab [1]{#1}%
\providecommand \enquote  [1]{``#1''}%
\providecommand \bibnamefont  [1]{#1}%
\providecommand \bibfnamefont [1]{#1}%
\providecommand \citenamefont [1]{#1}%
\providecommand \href@noop [0]{\@secondoftwo}%
\providecommand \href [0]{\begingroup \@sanitize@url \@href}%
\providecommand \@href[1]{\@@startlink{#1}\@@href}%
\providecommand \@@href[1]{\endgroup#1\@@endlink}%
\providecommand \@sanitize@url [0]{\catcode `\\12\catcode `\$12\catcode
  `\&12\catcode `\#12\catcode `\^12\catcode `\_12\catcode `\%12\relax}%
\providecommand \@@startlink[1]{}%
\providecommand \@@endlink[0]{}%
\providecommand \url  [0]{\begingroup\@sanitize@url \@url }%
\providecommand \@url [1]{\endgroup\@href {#1}{\urlprefix }}%
\providecommand \urlprefix  [0]{URL }%
\providecommand \Eprint [0]{\href }%
\providecommand \doibase [0]{https://doi.org/}%
\providecommand \selectlanguage [0]{\@gobble}%
\providecommand \bibinfo  [0]{\@secondoftwo}%
\providecommand \bibfield  [0]{\@secondoftwo}%
\providecommand \translation [1]{[#1]}%
\providecommand \BibitemOpen [0]{}%
\providecommand \bibitemStop [0]{}%
\providecommand \bibitemNoStop [0]{.\EOS\space}%
\providecommand \EOS [0]{\spacefactor3000\relax}%
\providecommand \BibitemShut  [1]{\csname bibitem#1\endcsname}%
\let\auto@bib@innerbib\@empty
\bibitem [{\citenamefont {Coussot}(2005)}]{Coussot:2005}%
  \BibitemOpen
  \bibfield  {author} {\bibinfo {author} {\bibfnamefont {P.}~\bibnamefont
  {Coussot}},\ }\href@noop {} {\emph {\bibinfo {title} {Rheometry of pastes,
  suspensions, and granular materials: applications in industry and
  environment}}}\ (\bibinfo  {publisher} {John Wiley \& Sons},\ \bibinfo {year}
  {2005})\BibitemShut {NoStop}%
\bibitem [{\citenamefont {Harnby}\ \emph {et~al.}(1997)\citenamefont {Harnby},
  \citenamefont {Edwards},\ and\ \citenamefont {Nienow}}]{Harnby_1997}%
  \BibitemOpen
  \bibfield  {author} {\bibinfo {author} {\bibfnamefont {N.}~\bibnamefont
  {Harnby}}, \bibinfo {author} {\bibfnamefont {M.~F.}\ \bibnamefont
  {Edwards}},\ and\ \bibinfo {author} {\bibfnamefont {A.~W.}\ \bibnamefont
  {Nienow}},\ }\href@noop {} {\emph {\bibinfo {title} {Mixing in the Process
  Industries: Second Edition}}}\ (\bibinfo  {publisher} {BH},\ \bibinfo {year}
  {1997})\BibitemShut {NoStop}%
\bibitem [{\citenamefont {Goldszal}\ and\ \citenamefont
  {Bousquet}(2001)}]{Goldszal:2001}%
  \BibitemOpen
  \bibfield  {author} {\bibinfo {author} {\bibfnamefont {A.}~\bibnamefont
  {Goldszal}}\ and\ \bibinfo {author} {\bibfnamefont {J.}~\bibnamefont
  {Bousquet}},\ }\bibfield  {title} {\bibinfo {title} {Wet agglomeration of
  powders: from physics toward process optimization},\ }\href@noop {}
  {\bibfield  {journal} {\bibinfo  {journal} {Powder Technol.}\ }\textbf
  {\bibinfo {volume} {117}},\ \bibinfo {pages} {221} (\bibinfo {year}
  {2001})}\BibitemShut {NoStop}%
\bibitem [{\citenamefont {Dhanalakshmi}\ \emph {et~al.}(2011)\citenamefont
  {Dhanalakshmi}, \citenamefont {Ghosal},\ and\ \citenamefont
  {Bhattacharya}}]{Dhanalakshmi:2011}%
  \BibitemOpen
  \bibfield  {author} {\bibinfo {author} {\bibfnamefont {K.}~\bibnamefont
  {Dhanalakshmi}}, \bibinfo {author} {\bibfnamefont {S.}~\bibnamefont
  {Ghosal}},\ and\ \bibinfo {author} {\bibfnamefont {S.}~\bibnamefont
  {Bhattacharya}},\ }\bibfield  {title} {\bibinfo {title} {Agglomeration of
  food powder and applications},\ }\href@noop {} {\bibfield  {journal}
  {\bibinfo  {journal} {Crit. Rev. Food Sci. Nutr.}\ }\textbf {\bibinfo
  {volume} {51}},\ \bibinfo {pages} {432} (\bibinfo {year} {2011})}\BibitemShut
  {NoStop}%
\bibitem [{\citenamefont {Hansuld}\ and\ \citenamefont
  {Briens}(2014)}]{Hansuld:2014}%
  \BibitemOpen
  \bibfield  {author} {\bibinfo {author} {\bibfnamefont {E.}~\bibnamefont
  {Hansuld}}\ and\ \bibinfo {author} {\bibfnamefont {L.}~\bibnamefont
  {Briens}},\ }\bibfield  {title} {\bibinfo {title} {A review of monitoring
  methods for pharmaceutical wet granulation},\ }\href@noop {} {\bibfield
  {journal} {\bibinfo  {journal} {Int. J. Pharm.}\ }\textbf {\bibinfo {volume}
  {472}},\ \bibinfo {pages} {192} (\bibinfo {year} {2014})}\BibitemShut
  {NoStop}%
\bibitem [{\citenamefont {Rondet}\ \emph {et~al.}(2008)\citenamefont {Rondet},
  \citenamefont {Delalonde}, \citenamefont {Ruiz},\ and\ \citenamefont
  {Desfours}}]{Rondet:2008}%
  \BibitemOpen
  \bibfield  {author} {\bibinfo {author} {\bibfnamefont {E.}~\bibnamefont
  {Rondet}}, \bibinfo {author} {\bibfnamefont {M.}~\bibnamefont {Delalonde}},
  \bibinfo {author} {\bibfnamefont {T.}~\bibnamefont {Ruiz}},\ and\ \bibinfo
  {author} {\bibfnamefont {J.-P.}\ \bibnamefont {Desfours}},\ }\bibfield
  {title} {\bibinfo {title} {Hydro-textural and dimensional approach for
  characterising wet granular media agglomerated by kneading},\ }\href@noop {}
  {\bibfield  {journal} {\bibinfo  {journal} {Chem. Eng. Res. Des.}\ }\textbf
  {\bibinfo {volume} {86}},\ \bibinfo {pages} {560} (\bibinfo {year}
  {2008})}\BibitemShut {NoStop}%
\bibitem [{\citenamefont {Rondet}\ \emph
  {et~al.}(2009{\natexlab{a}})\citenamefont {Rondet}, \citenamefont
  {Delalonde}, \citenamefont {Ruiz},\ and\ \citenamefont
  {Desfours}}]{Rondet:2009a}%
  \BibitemOpen
  \bibfield  {author} {\bibinfo {author} {\bibfnamefont {E.}~\bibnamefont
  {Rondet}}, \bibinfo {author} {\bibfnamefont {M.}~\bibnamefont {Delalonde}},
  \bibinfo {author} {\bibfnamefont {T.}~\bibnamefont {Ruiz}},\ and\ \bibinfo
  {author} {\bibfnamefont {J.-P.}\ \bibnamefont {Desfours}},\ }\bibfield
  {title} {\bibinfo {title} {Identification of granular compactness during the
  kneading of a humidified cohesive powder},\ }\href@noop {} {\bibfield
  {journal} {\bibinfo  {journal} {Powder Technol.}\ }\textbf {\bibinfo {volume}
  {191}},\ \bibinfo {pages} {7} (\bibinfo {year}
  {2009}{\natexlab{a}})}\BibitemShut {NoStop}%
\bibitem [{\citenamefont {Paul}\ \emph {et~al.}(2003)\citenamefont {Paul},
  \citenamefont {Atiemo-Obeng},\ and\ \citenamefont {Kresta}}]{Paul_2003}%
  \BibitemOpen
  \bibfield  {author} {\bibinfo {author} {\bibfnamefont {E.~L.}\ \bibnamefont
  {Paul}}, \bibinfo {author} {\bibfnamefont {V.~A.}\ \bibnamefont
  {Atiemo-Obeng}},\ and\ \bibinfo {author} {\bibfnamefont {S.~M.}\ \bibnamefont
  {Kresta}},\ }\href@noop {} {\emph {\bibinfo {title} {Handbook of Industrial
  Mixing: Science and Practice}}}\ (\bibinfo  {publisher} {John Wiley \& Sons,
  Inc.},\ \bibinfo {year} {2003})\BibitemShut {NoStop}%
\bibitem [{\citenamefont {Cappelli}\ \emph {et~al.}(2020)\citenamefont
  {Cappelli}, \citenamefont {Bettaccini},\ and\ \citenamefont
  {Cini}}]{Campelli_2020}%
  \BibitemOpen
  \bibfield  {author} {\bibinfo {author} {\bibfnamefont {A.}~\bibnamefont
  {Cappelli}}, \bibinfo {author} {\bibfnamefont {L.}~\bibnamefont
  {Bettaccini}},\ and\ \bibinfo {author} {\bibfnamefont {E.}~\bibnamefont
  {Cini}},\ }\bibfield  {title} {\bibinfo {title} {The kneading process: A
  systematic review of the effects on dough rheology and resulting bread
  characteristics, including improvement strategies},\ }\href@noop {}
  {\bibfield  {journal} {\bibinfo  {journal} {Trends Food Sci. Technol.}\
  }\textbf {\bibinfo {volume} {104}},\ \bibinfo {pages} {91} (\bibinfo {year}
  {2020})}\BibitemShut {NoStop}%
\bibitem [{\citenamefont {Adragna}(2006)}]{Adragna_2006}%
  \BibitemOpen
  \bibfield  {author} {\bibinfo {author} {\bibfnamefont {L.}~\bibnamefont
  {Adragna}},\ }\emph {\bibinfo {title} {Mise en œuvre réactive des
  polymères étude et modélisation de la dispersion en mélangeur interne
  d'un liquide peu visqueux dans un polymère fondu}},\ \href@noop {} {\bibinfo
  {type} {{Ph.D.} thesis}},\ \bibinfo  {school} {Université Claude Bernard -
  Lyon I} (\bibinfo {year} {2006})\BibitemShut {NoStop}%
\bibitem [{\citenamefont {Le~Lan}(1983)}]{LeLan_1983}%
  \BibitemOpen
  \bibfield  {author} {\bibinfo {author} {\bibfnamefont {A.}~\bibnamefont
  {Le~Lan}},\ }\bibfield  {title} {\bibinfo {title} {Mélange de pâtes},\
  }\href@noop {} {\bibfield  {journal} {\bibinfo  {journal} {Techniques de
  l'Ingénieur}\ } (\bibinfo {year} {1983})}\BibitemShut {NoStop}%
\bibitem [{\citenamefont {Albright}\ \emph {et~al.}(2009)\citenamefont
  {Albright}, \citenamefont {Trinh},\ and\ \citenamefont
  {Harvey}}]{Albright_2009}%
  \BibitemOpen
  \bibfield  {author} {\bibinfo {author} {\bibfnamefont {L.~F.}\ \bibnamefont
  {Albright}}, \bibinfo {author} {\bibfnamefont {S.}~\bibnamefont {Trinh}},\
  and\ \bibinfo {author} {\bibfnamefont {A.~H.}\ \bibnamefont {Harvey}},\
  }\href@noop {} {\emph {\bibinfo {title} {ALBRIGHT'S chemical engineering
  handbook}}}\ (\bibinfo  {publisher} {CRC Press},\ \bibinfo {year}
  {2009})\BibitemShut {NoStop}%
\bibitem [{\citenamefont {Seem}\ \emph {et~al.}(2015)\citenamefont {Seem},
  \citenamefont {Rowson}, \citenamefont {Ingram}, \citenamefont {Huang},
  \citenamefont {Yu}, \citenamefont {Matas}, \citenamefont {Gabbott},\ and\
  \citenamefont {Reynolds}}]{Seem:2015}%
  \BibitemOpen
  \bibfield  {author} {\bibinfo {author} {\bibfnamefont {T.~C.}\ \bibnamefont
  {Seem}}, \bibinfo {author} {\bibfnamefont {N.~A.}\ \bibnamefont {Rowson}},
  \bibinfo {author} {\bibfnamefont {A.}~\bibnamefont {Ingram}}, \bibinfo
  {author} {\bibfnamefont {Z.}~\bibnamefont {Huang}}, \bibinfo {author}
  {\bibfnamefont {S.}~\bibnamefont {Yu}}, \bibinfo {author} {\bibfnamefont
  {M.~d.}\ \bibnamefont {Matas}}, \bibinfo {author} {\bibfnamefont
  {I.}~\bibnamefont {Gabbott}},\ and\ \bibinfo {author} {\bibfnamefont {G.~K.}\
  \bibnamefont {Reynolds}},\ }\bibfield  {title} {\bibinfo {title} {Twin screw
  granulation -- {A} literature review},\ }\href@noop {} {\bibfield  {journal}
  {\bibinfo  {journal} {Powder Technol.}\ }\textbf {\bibinfo {volume} {276}},\
  \bibinfo {pages} {89} (\bibinfo {year} {2015})}\BibitemShut {NoStop}%
\bibitem [{\citenamefont {Lloyd}(2011)}]{Lloyd_2011}%
  \BibitemOpen
  \bibfield  {author} {\bibinfo {author} {\bibfnamefont {L.}~\bibnamefont
  {Lloyd}},\ }\href@noop {} {\emph {\bibinfo {title} {Handbook of Industrial
  Catalysts}}}\ (\bibinfo  {publisher} {Springer Science \& Business Media},\
  \bibinfo {year} {2011})\BibitemShut {NoStop}%
\bibitem [{\citenamefont {Dubois}\ and\ \citenamefont
  {Fujieda}(1995)}]{Dubois:1995}%
  \BibitemOpen
  \bibfield  {author} {\bibinfo {author} {\bibfnamefont {J.-L.}\ \bibnamefont
  {Dubois}}\ and\ \bibinfo {author} {\bibfnamefont {S.}~\bibnamefont
  {Fujieda}},\ }\bibfield  {title} {\bibinfo {title} {Preparation of
  boron-containing and alumina supports by kneading},\ }in\ \href@noop {}
  {\emph {\bibinfo {booktitle} {Stud. Surf. Sci. Catal.}}},\ Vol.~\bibinfo
  {volume} {91}\ (\bibinfo {year} {1995})\ pp.\ \bibinfo {pages}
  {833--842}\BibitemShut {NoStop}%
\bibitem [{\citenamefont {Vatutina}\ \emph {et~al.}(2016)\citenamefont
  {Vatutina}, \citenamefont {Klimov}, \citenamefont {Nadeina}, \citenamefont
  {Danilovaand}, \citenamefont {Gerasimov}, \citenamefont {Prosvirin},\ and\
  \citenamefont {Noskov}}]{Vatutina_2016}%
  \BibitemOpen
  \bibfield  {author} {\bibinfo {author} {\bibfnamefont {Y.}~\bibnamefont
  {Vatutina}}, \bibinfo {author} {\bibfnamefont {O.}~\bibnamefont {Klimov}},
  \bibinfo {author} {\bibfnamefont {K.}~\bibnamefont {Nadeina}}, \bibinfo
  {author} {\bibfnamefont {I.}~\bibnamefont {Danilovaand}}, \bibinfo {author}
  {\bibfnamefont {E.}~\bibnamefont {Gerasimov}}, \bibinfo {author}
  {\bibfnamefont {I.}~\bibnamefont {Prosvirin}},\ and\ \bibinfo {author}
  {\bibfnamefont {A.}~\bibnamefont {Noskov}},\ }\bibfield  {title} {\bibinfo
  {title} {Influence of boron addition to alumina support by kneading on
  morphology and activity of hds catalysts},\ }\href@noop {} {\bibfield
  {journal} {\bibinfo  {journal} {Appl. Catal. B}\ }\textbf {\bibinfo {volume}
  {199}},\ \bibinfo {pages} {23} (\bibinfo {year} {2016})}\BibitemShut
  {NoStop}%
\bibitem [{\citenamefont {Euzen}\ \emph {et~al.}(2008)\citenamefont {Euzen},
  \citenamefont {Raybaud}, \citenamefont {Krokidis}, \citenamefont {Toulhoat},
  \citenamefont {Le~Loarer}, \citenamefont {Jolivet},\ and\ \citenamefont
  {Froidefond}}]{Euzen_2008}%
  \BibitemOpen
  \bibfield  {author} {\bibinfo {author} {\bibfnamefont {P.}~\bibnamefont
  {Euzen}}, \bibinfo {author} {\bibfnamefont {P.}~\bibnamefont {Raybaud}},
  \bibinfo {author} {\bibfnamefont {X.}~\bibnamefont {Krokidis}}, \bibinfo
  {author} {\bibfnamefont {H.}~\bibnamefont {Toulhoat}}, \bibinfo {author}
  {\bibfnamefont {J.-L.}\ \bibnamefont {Le~Loarer}}, \bibinfo {author}
  {\bibfnamefont {J.-P.}\ \bibnamefont {Jolivet}},\ and\ \bibinfo {author}
  {\bibfnamefont {C.}~\bibnamefont {Froidefond}},\ }\href@noop {} {\emph
  {\bibinfo {title} {Handbook of porous solids: Alumina}}}\ (\bibinfo
  {publisher} {Wiley-VCH},\ \bibinfo {year} {2008})\BibitemShut {NoStop}%
\bibitem [{\citenamefont {Klaewkla}\ \emph {et~al.}(2011)\citenamefont
  {Klaewkla}, \citenamefont {Arend},\ and\ \citenamefont
  {Hoelderich}}]{Klaewkla:2011}%
  \BibitemOpen
  \bibfield  {author} {\bibinfo {author} {\bibfnamefont {R.}~\bibnamefont
  {Klaewkla}}, \bibinfo {author} {\bibfnamefont {M.}~\bibnamefont {Arend}},\
  and\ \bibinfo {author} {\bibfnamefont {W.}~\bibnamefont {Hoelderich}},\
  }\href@noop {} {\emph {\bibinfo {title} {Mass Transfer - Advanced Aspects}}}\
  (\bibinfo  {publisher} {InTech},\ \bibinfo {year} {2011})\BibitemShut
  {NoStop}%
\bibitem [{\citenamefont {Hart}\ and\ \citenamefont {Lense}(1990)}]{Hart_1990}%
  \BibitemOpen
  \bibfield  {author} {\bibinfo {author} {\bibfnamefont {L.~D.}\ \bibnamefont
  {Hart}}\ and\ \bibinfo {author} {\bibfnamefont {E.}~\bibnamefont {Lense}},\
  }\href@noop {} {\emph {\bibinfo {title} {Alumina Chemicals: Science and
  Technology Handbook}}}\ (\bibinfo  {publisher} {John Wiley \& Sons},\
  \bibinfo {year} {1990})\BibitemShut {NoStop}%
\bibitem [{\citenamefont {Trimm}(1986)}]{Trimm_1986}%
  \BibitemOpen
  \bibfield  {author} {\bibinfo {author} {\bibfnamefont {A.}~\bibnamefont
  {Trimm}, \bibfnamefont {D.~L.and~Stanislaus}},\ }\bibfield  {title} {\bibinfo
  {title} {The control of pore size in alumina catalyst supports: A review},\
  }\href@noop {} {\bibfield  {journal} {\bibinfo  {journal} {Appl. Catal.}\
  }\textbf {\bibinfo {volume} {21}},\ \bibinfo {pages} {215–238} (\bibinfo
  {year} {1986})}\BibitemShut {NoStop}%
\bibitem [{\citenamefont {Kolitcheff}\ \emph {et~al.}(2017)\citenamefont
  {Kolitcheff}, \citenamefont {Jolimaitre}, \citenamefont {Hugon},
  \citenamefont {Verstraete}, \citenamefont {Carrette},\ and\ \citenamefont
  {Tayakout-Fayolle}}]{Kolitcheff:2017}%
  \BibitemOpen
  \bibfield  {author} {\bibinfo {author} {\bibfnamefont {S.}~\bibnamefont
  {Kolitcheff}}, \bibinfo {author} {\bibfnamefont {E.}~\bibnamefont
  {Jolimaitre}}, \bibinfo {author} {\bibfnamefont {A.}~\bibnamefont {Hugon}},
  \bibinfo {author} {\bibfnamefont {J.}~\bibnamefont {Verstraete}}, \bibinfo
  {author} {\bibfnamefont {P.-L.}\ \bibnamefont {Carrette}},\ and\ \bibinfo
  {author} {\bibfnamefont {M.}~\bibnamefont {Tayakout-Fayolle}},\ }\bibfield
  {title} {\bibinfo {title} {Tortuosity of mesoporous alumina catalyst
  supports: Influence of the pore network organization},\ }\href@noop {}
  {\bibfield  {journal} {\bibinfo  {journal} {Microporous and Mesoporous
  Mater.}\ }\textbf {\bibinfo {volume} {248}},\ \bibinfo {pages} {91} (\bibinfo
  {year} {2017})}\BibitemShut {NoStop}%
\bibitem [{\citenamefont {Pourcel}\ \emph {et~al.}(2007)\citenamefont
  {Pourcel}, \citenamefont {Jomaa}, \citenamefont {Puiggali},\ and\
  \citenamefont {Rouleau}}]{Pourcel_2007}%
  \BibitemOpen
  \bibfield  {author} {\bibinfo {author} {\bibfnamefont {F.}~\bibnamefont
  {Pourcel}}, \bibinfo {author} {\bibfnamefont {W.}~\bibnamefont {Jomaa}},
  \bibinfo {author} {\bibfnamefont {J.-R.}\ \bibnamefont {Puiggali}},\ and\
  \bibinfo {author} {\bibfnamefont {L.}~\bibnamefont {Rouleau}},\ }\bibfield
  {title} {\bibinfo {title} {Criterion for crack initiation during drying:
  Alumina porous ceramic strength improvement},\ }\href@noop {} {\bibfield
  {journal} {\bibinfo  {journal} {Powder Technol.}\ }\textbf {\bibinfo {volume}
  {172}},\ \bibinfo {pages} {120} (\bibinfo {year} {2007})}\BibitemShut
  {NoStop}%
\bibitem [{\citenamefont {Karouia}\ \emph
  {et~al.}(2013{\natexlab{a}})\citenamefont {Karouia}, \citenamefont
  {Boualleg}, \citenamefont {Digne},\ and\ \citenamefont
  {Alphonse}}]{Karouia_13}%
  \BibitemOpen
  \bibfield  {author} {\bibinfo {author} {\bibfnamefont {F.}~\bibnamefont
  {Karouia}}, \bibinfo {author} {\bibfnamefont {M.}~\bibnamefont {Boualleg}},
  \bibinfo {author} {\bibfnamefont {M.}~\bibnamefont {Digne}},\ and\ \bibinfo
  {author} {\bibfnamefont {P.}~\bibnamefont {Alphonse}},\ }\bibfield  {title}
  {\bibinfo {title} {Control of the textural properties of nanocrystalline
  boehmite ($\gamma$-alooh) regarding its peptization ability},\ }\href@noop {}
  {\bibfield  {journal} {\bibinfo  {journal} {Powder Technol.}\ }\textbf
  {\bibinfo {volume} {237}},\ \bibinfo {pages} {602–609} (\bibinfo {year}
  {2013}{\natexlab{a}})}\BibitemShut {NoStop}%
\bibitem [{\citenamefont {Speyer}\ \emph {et~al.}(2020)\citenamefont {Speyer},
  \citenamefont {Humbert}, \citenamefont {Bizien}, \citenamefont {Lecocq},\
  and\ \citenamefont {Hugon}}]{Speyer_2020}%
  \BibitemOpen
  \bibfield  {author} {\bibinfo {author} {\bibfnamefont {L.}~\bibnamefont
  {Speyer}}, \bibinfo {author} {\bibfnamefont {S.}~\bibnamefont {Humbert}},
  \bibinfo {author} {\bibfnamefont {T.}~\bibnamefont {Bizien}}, \bibinfo
  {author} {\bibfnamefont {V.}~\bibnamefont {Lecocq}},\ and\ \bibinfo {author}
  {\bibfnamefont {A.}~\bibnamefont {Hugon}},\ }\bibfield  {title} {\bibinfo
  {title} {Peptization of boehmites with different peptization index: An
  electron microscopy and synchrotron small-angle {X}-ray scattering study},\
  }\href@noop {} {\bibfield  {journal} {\bibinfo  {journal} {Colloids Surf.}\
  }\textbf {\bibinfo {volume} {603}},\ \bibinfo {pages} {125175} (\bibinfo
  {year} {2020})}\BibitemShut {NoStop}%
\bibitem [{\citenamefont {Ramsay}\ \emph {et~al.}(1978)\citenamefont {Ramsay},
  \citenamefont {Daish},\ and\ \citenamefont {Wright}}]{Ramsay:1978}%
  \BibitemOpen
  \bibfield  {author} {\bibinfo {author} {\bibfnamefont {J.}~\bibnamefont
  {Ramsay}}, \bibinfo {author} {\bibfnamefont {S.}~\bibnamefont {Daish}},\ and\
  \bibinfo {author} {\bibfnamefont {C.}~\bibnamefont {Wright}},\ }\bibfield
  {title} {\bibinfo {title} {Structure and stability of concentrated boehmite
  sols},\ }\href@noop {} {\bibfield  {journal} {\bibinfo  {journal} {Faraday
  Discuss.}\ }\textbf {\bibinfo {volume} {65}},\ \bibinfo {pages} {65}
  (\bibinfo {year} {1978})}\BibitemShut {NoStop}%
\bibitem [{\citenamefont {Fauchadour}\ \emph {et~al.}(2002)\citenamefont
  {Fauchadour}, \citenamefont {Kolenda}, \citenamefont {Rouleau}, \citenamefont
  {Barré},\ and\ \citenamefont {Normand}}]{Fauchadour_2002}%
  \BibitemOpen
  \bibfield  {author} {\bibinfo {author} {\bibfnamefont {D.}~\bibnamefont
  {Fauchadour}}, \bibinfo {author} {\bibfnamefont {F.}~\bibnamefont {Kolenda}},
  \bibinfo {author} {\bibfnamefont {L.}~\bibnamefont {Rouleau}}, \bibinfo
  {author} {\bibfnamefont {L.}~\bibnamefont {Barré}},\ and\ \bibinfo {author}
  {\bibfnamefont {L.}~\bibnamefont {Normand}},\ }\bibfield  {title} {\bibinfo
  {title} {Peptization mechanisms of boehmite used as precursors for
  catalysts},\ }\href@noop {} {\bibfield  {journal} {\bibinfo  {journal} {Ind.
  Eng. Chem. Res.}\ }\textbf {\bibinfo {volume} {143}},\ \bibinfo {pages}
  {453–461} (\bibinfo {year} {2002})}\BibitemShut {NoStop}%
\bibitem [{\citenamefont {Zheng}\ \emph {et~al.}(2014)\citenamefont {Zheng},
  \citenamefont {Song}, \citenamefont {Xu}, \citenamefont {He}, \citenamefont
  {Wang},\ and\ \citenamefont {Yan}}]{Zheng_2014}%
  \BibitemOpen
  \bibfield  {author} {\bibinfo {author} {\bibfnamefont {Y.}~\bibnamefont
  {Zheng}}, \bibinfo {author} {\bibfnamefont {J.}~\bibnamefont {Song}},
  \bibinfo {author} {\bibfnamefont {X.}~\bibnamefont {Xu}}, \bibinfo {author}
  {\bibfnamefont {M.}~\bibnamefont {He}}, \bibinfo {author} {\bibfnamefont
  {Q.}~\bibnamefont {Wang}},\ and\ \bibinfo {author} {\bibfnamefont
  {L.}~\bibnamefont {Yan}},\ }\bibfield  {title} {\bibinfo {title} {Peptization
  mechanism of boehmite and its effect on the preparation of a fluid catalytic
  cracking catalyst},\ }\href@noop {} {\bibfield  {journal} {\bibinfo
  {journal} {Ind. Eng. Chem. Res.}\ }\textbf {\bibinfo {volume} {53}},\
  \bibinfo {pages} {10029–10034} (\bibinfo {year} {2014})}\BibitemShut
  {NoStop}%
\bibitem [{\citenamefont {Danner}\ and\ \citenamefont
  {Unger}(1987)}]{Danner:1987}%
  \BibitemOpen
  \bibfield  {author} {\bibinfo {author} {\bibfnamefont {A.}~\bibnamefont
  {Danner}}\ and\ \bibinfo {author} {\bibfnamefont {K.~K.}\ \bibnamefont
  {Unger}},\ }\bibfield  {title} {\bibinfo {title} {The relevance of kneading
  and extrusion parameters in the manufacture of active porous aluminas from
  pseudoboehmites},\ }\href@noop {} {\bibfield  {journal} {\bibinfo  {journal}
  {Stud. Surf. Sci. Catal.}\ }\textbf {\bibinfo {volume} {31}},\ \bibinfo
  {pages} {343} (\bibinfo {year} {1987})}\BibitemShut {NoStop}%
\bibitem [{\citenamefont {Studart}\ \emph {et~al.}(2006)\citenamefont
  {Studart}, \citenamefont {Gonzenbach}, \citenamefont {Tervoort},\ and\
  \citenamefont {Gauckler}}]{Studart:2006}%
  \BibitemOpen
  \bibfield  {author} {\bibinfo {author} {\bibfnamefont {A.~R.}\ \bibnamefont
  {Studart}}, \bibinfo {author} {\bibfnamefont {U.~T.}\ \bibnamefont
  {Gonzenbach}}, \bibinfo {author} {\bibfnamefont {E.}~\bibnamefont
  {Tervoort}},\ and\ \bibinfo {author} {\bibfnamefont {L.~J.}\ \bibnamefont
  {Gauckler}},\ }\bibfield  {title} {\bibinfo {title} {Processing routes to
  macroporous ceramics: a review},\ }\href@noop {} {\bibfield  {journal}
  {\bibinfo  {journal} {J. Am. Ceram. Soc.}\ }\textbf {\bibinfo {volume}
  {89}},\ \bibinfo {pages} {1771} (\bibinfo {year} {2006})}\BibitemShut
  {NoStop}%
\bibitem [{\citenamefont {M{\'a}rquez-Alvarez}\ \emph
  {et~al.}(2008)\citenamefont {M{\'a}rquez-Alvarez}, \citenamefont
  {{\v{Z}}ilkov{\'a}}, \citenamefont {P{\'e}rez-Pariente},\ and\ \citenamefont
  {{\v{C}}ejka}}]{MarquezAlvarez:2008}%
  \BibitemOpen
  \bibfield  {author} {\bibinfo {author} {\bibfnamefont {C.}~\bibnamefont
  {M{\'a}rquez-Alvarez}}, \bibinfo {author} {\bibfnamefont {N.}~\bibnamefont
  {{\v{Z}}ilkov{\'a}}}, \bibinfo {author} {\bibfnamefont {J.}~\bibnamefont
  {P{\'e}rez-Pariente}},\ and\ \bibinfo {author} {\bibfnamefont
  {J.}~\bibnamefont {{\v{C}}ejka}},\ }\bibfield  {title} {\bibinfo {title}
  {Synthesis, characterization and catalytic applications of organized
  mesoporous aluminas},\ }\href@noop {} {\bibfield  {journal} {\bibinfo
  {journal} {Catal. Rev.}\ }\textbf {\bibinfo {volume} {50}},\ \bibinfo {pages}
  {222} (\bibinfo {year} {2008})}\BibitemShut {NoStop}%
\bibitem [{\citenamefont {Walendziewski}\ and\ \citenamefont
  {Trawczyfiski}(1994)}]{Walendziewski:1994}%
  \BibitemOpen
  \bibfield  {author} {\bibinfo {author} {\bibfnamefont {J.}~\bibnamefont
  {Walendziewski}}\ and\ \bibinfo {author} {\bibfnamefont {J.}~\bibnamefont
  {Trawczyfiski}},\ }\bibfield  {title} {\bibinfo {title} {Influence of the
  forming method on the pore structure of alumina supports},\ }\href@noop {}
  {\bibfield  {journal} {\bibinfo  {journal} {Appl. Catal. A: Gen.}\ }\textbf
  {\bibinfo {volume} {119}},\ \bibinfo {pages} {45} (\bibinfo {year}
  {1994})}\BibitemShut {NoStop}%
\bibitem [{\citenamefont {Landers}\ \emph {et~al.}(2010)\citenamefont
  {Landers}, \citenamefont {Devadas}, \citenamefont {Neimark}, \citenamefont
  {Timken}, \citenamefont {Ojo},\ and\ \citenamefont {Chester}}]{Landers_2010}%
  \BibitemOpen
  \bibfield  {author} {\bibinfo {author} {\bibfnamefont {J.}~\bibnamefont
  {Landers}}, \bibinfo {author} {\bibfnamefont {M.}~\bibnamefont {Devadas}},
  \bibinfo {author} {\bibfnamefont {A.~V.}\ \bibnamefont {Neimark}}, \bibinfo
  {author} {\bibfnamefont {H.-K.}\ \bibnamefont {Timken}}, \bibinfo {author}
  {\bibfnamefont {A.}~\bibnamefont {Ojo}},\ and\ \bibinfo {author}
  {\bibfnamefont {A.~W.}\ \bibnamefont {Chester}},\ }\bibfield  {title}
  {\bibinfo {title} {Effect ofmixing on the pore structure ofalumina
  extrudates},\ }\href@noop {} {\bibfield  {journal} {\bibinfo  {journal}
  {Part. Part. Syst. Charact.}\ }\textbf {\bibinfo {volume} {27}},\ \bibinfo
  {pages} {42} (\bibinfo {year} {2010})}\BibitemShut {NoStop}%
\bibitem [{\citenamefont {Sudreau}\ \emph {et~al.}(2023)\citenamefont
  {Sudreau}, \citenamefont {Servel}, \citenamefont {Freyssingeas},
  \citenamefont {Liénard}, \citenamefont {Karpati}, \citenamefont {Parola},
  \citenamefont {Jaurand}, \citenamefont {Dugas}, \citenamefont {Matthews},
  \citenamefont {Gibaud}, \citenamefont {Divoux},\ and\ \citenamefont
  {Manneville}}]{Sudreau_2023}%
  \BibitemOpen
  \bibfield  {author} {\bibinfo {author} {\bibfnamefont {I.}~\bibnamefont
  {Sudreau}}, \bibinfo {author} {\bibfnamefont {M.}~\bibnamefont {Servel}},
  \bibinfo {author} {\bibfnamefont {E.}~\bibnamefont {Freyssingeas}}, \bibinfo
  {author} {\bibfnamefont {F.}~\bibnamefont {Liénard}}, \bibinfo {author}
  {\bibfnamefont {S.}~\bibnamefont {Karpati}}, \bibinfo {author} {\bibfnamefont
  {S.}~\bibnamefont {Parola}}, \bibinfo {author} {\bibfnamefont
  {X.}~\bibnamefont {Jaurand}}, \bibinfo {author} {\bibfnamefont {P.-Y.}\
  \bibnamefont {Dugas}}, \bibinfo {author} {\bibfnamefont {L.}~\bibnamefont
  {Matthews}}, \bibinfo {author} {\bibfnamefont {T.}~\bibnamefont {Gibaud}},
  \bibinfo {author} {\bibfnamefont {T.}~\bibnamefont {Divoux}},\ and\ \bibinfo
  {author} {\bibfnamefont {S.}~\bibnamefont {Manneville}},\ }\bibfield  {title}
  {\bibinfo {title} {Shear-induced stiffening in boehmite gels: A
  rheo-{X}-ray-scattering study},\ }\href@noop {} {\bibfield  {journal}
  {\bibinfo  {journal} {Phys. Rev. Materials}\ }\textbf {\bibinfo {volume}
  {7}},\ \bibinfo {pages} {115603} (\bibinfo {year} {2023})}\BibitemShut
  {NoStop}%
\bibitem [{\citenamefont {Gallois}(2016)}]{Gallois_2016}%
  \BibitemOpen
  \bibfield  {author} {\bibinfo {author} {\bibfnamefont {C.}~\bibnamefont
  {Gallois}},\ }\emph {\bibinfo {title} {Etude des propriétés
  physico-chimiques de suspensions de boehmite. Application aux supports
  catalytiques}},\ \href@noop {} {\bibinfo {type} {{Ph.D.} thesis}},\ \bibinfo
  {school} {Université Pierre et Marie Curie - Paris VI} (\bibinfo {year}
  {2016})\BibitemShut {NoStop}%
\bibitem [{\citenamefont {Smith}(2014)}]{Morin_2014}%
  \BibitemOpen
  \bibfield  {author} {\bibinfo {author} {\bibfnamefont {S.~M.}\ \bibnamefont
  {Smith}},\ }\emph {\bibinfo {title} {Péeparation d'alumine à porosité
  contrôlée : étude de l'interaction de la boehmite dans des solvants et des
  propriétés fonctionnelles des matériaux résultants.}},\ \href@noop {}
  {\bibinfo {type} {{Ph.D.} thesis}},\ \bibinfo  {school} {Université Pierre
  et Marie Curie - Paris VI} (\bibinfo {year} {2014})\BibitemShut {NoStop}%
\bibitem [{\citenamefont {Schlumberger}\ and\ \citenamefont
  {Thommes}(2021)}]{Schlumberger_2021}%
  \BibitemOpen
  \bibfield  {author} {\bibinfo {author} {\bibfnamefont {C.}~\bibnamefont
  {Schlumberger}}\ and\ \bibinfo {author} {\bibfnamefont {M.}~\bibnamefont
  {Thommes}},\ }\bibfield  {title} {\bibinfo {title} {Characterization of
  hierarchically ordered porous materials by physisorption and mercury
  porosimetry—a tutorial review},\ }\href@noop {} {\bibfield  {journal}
  {\bibinfo  {journal} {Adv. Mater. Interfaces}\ }\textbf {\bibinfo {volume}
  {8}},\ \bibinfo {pages} {2002181} (\bibinfo {year} {2021})}\BibitemShut
  {NoStop}%
\bibitem [{\citenamefont {Rouquerol}\ \emph {et~al.}(1994)\citenamefont
  {Rouquerol}, \citenamefont {Avnir}, \citenamefont {Fairbridge}, \citenamefont
  {Everett}, \citenamefont {Haynes}, \citenamefont {Pernicone}, \citenamefont
  {Ramsay}, \citenamefont {Sing},\ and\ \citenamefont {Unger}}]{IUPAC_1994}%
  \BibitemOpen
  \bibfield  {author} {\bibinfo {author} {\bibfnamefont {J.}~\bibnamefont
  {Rouquerol}}, \bibinfo {author} {\bibfnamefont {D.}~\bibnamefont {Avnir}},
  \bibinfo {author} {\bibfnamefont {C.~W.}\ \bibnamefont {Fairbridge}},
  \bibinfo {author} {\bibfnamefont {D.~H.}\ \bibnamefont {Everett}}, \bibinfo
  {author} {\bibfnamefont {J.}~\bibnamefont {Haynes}}, \bibinfo {author}
  {\bibfnamefont {N.}~\bibnamefont {Pernicone}}, \bibinfo {author}
  {\bibfnamefont {J.~D.}\ \bibnamefont {Ramsay}}, \bibinfo {author}
  {\bibfnamefont {K.~S.~W.}\ \bibnamefont {Sing}},\ and\ \bibinfo {author}
  {\bibfnamefont {K.~K.}\ \bibnamefont {Unger}},\ }\bibfield  {title} {\bibinfo
  {title} {Recommendations for the characterization of porous solids (technical
  report)},\ }\href@noop {} {\bibfield  {journal} {\bibinfo  {journal} {Pure
  Appl. Chem.}\ }\textbf {\bibinfo {volume} {66}},\ \bibinfo {pages} {1739}
  (\bibinfo {year} {1994})}\BibitemShut {NoStop}%
\bibitem [{\citenamefont {Bogner}\ \emph {et~al.}(2020)\citenamefont {Bogner},
  \citenamefont {Schatz}, \citenamefont {Dehn},\ and\ \citenamefont
  {M{\"u}ller}}]{Bogner:2020}%
  \BibitemOpen
  \bibfield  {author} {\bibinfo {author} {\bibfnamefont {A.}~\bibnamefont
  {Bogner}}, \bibinfo {author} {\bibfnamefont {J.}~\bibnamefont {Schatz}},
  \bibinfo {author} {\bibfnamefont {F.}~\bibnamefont {Dehn}},\ and\ \bibinfo
  {author} {\bibfnamefont {H.~S.}\ \bibnamefont {M{\"u}ller}},\ }\bibfield
  {title} {\bibinfo {title} {Influence of drying on the microstructure of
  hardened cement paste: A mercury intrusion porosimetry, nitrogen sorption and
  saxs study},\ }\href@noop {} {\bibfield  {journal} {\bibinfo  {journal} {J.
  Adv. Concr. Technol.}\ }\textbf {\bibinfo {volume} {18}},\ \bibinfo {pages}
  {83} (\bibinfo {year} {2020})}\BibitemShut {NoStop}%
\bibitem [{\citenamefont {Brunauer}\ \emph {et~al.}(1938)\citenamefont
  {Brunauer}, \citenamefont {Deming}, \citenamefont {Deming},\ and\
  \citenamefont {Teller}}]{BET_1938}%
  \BibitemOpen
  \bibfield  {author} {\bibinfo {author} {\bibfnamefont {S.}~\bibnamefont
  {Brunauer}}, \bibinfo {author} {\bibfnamefont {L.}~\bibnamefont {Deming}},
  \bibinfo {author} {\bibfnamefont {E.}~\bibnamefont {Deming}},\ and\ \bibinfo
  {author} {\bibfnamefont {E.}~\bibnamefont {Teller}},\ }\bibfield  {title}
  {\bibinfo {title} {Adsorption of gases in multimolecular layers},\
  }\href@noop {} {\bibfield  {journal} {\bibinfo  {journal} {J. Am. Chem.
  Soc.}\ }\textbf {\bibinfo {volume} {60}},\ \bibinfo {pages} {309} (\bibinfo
  {year} {1938})}\BibitemShut {NoStop}%
\bibitem [{\citenamefont {Barrett}\ \emph {et~al.}(1951)\citenamefont
  {Barrett}, \citenamefont {Joyner},\ and\ \citenamefont {Halenda}}]{BJH_1951}%
  \BibitemOpen
  \bibfield  {author} {\bibinfo {author} {\bibfnamefont {E.}~\bibnamefont
  {Barrett}}, \bibinfo {author} {\bibfnamefont {L.}~\bibnamefont {Joyner}},\
  and\ \bibinfo {author} {\bibfnamefont {P.}~\bibnamefont {Halenda}},\
  }\bibfield  {title} {\bibinfo {title} {The determination of pore volume and
  area distributions in porous substances. {I.} {C}omputations from nitrogen
  isotherms},\ }\href@noop {} {\bibfield  {journal} {\bibinfo  {journal} {J.
  Am. Chem. Soc.}\ }\textbf {\bibinfo {volume} {73}},\ \bibinfo {pages}
  {373–380} (\bibinfo {year} {1951})}\BibitemShut {NoStop}%
\bibitem [{\citenamefont {Sing}(1985)}]{IUPAC_1985}%
  \BibitemOpen
  \bibfield  {author} {\bibinfo {author} {\bibfnamefont {K.~S.}\ \bibnamefont
  {Sing}},\ }\bibfield  {title} {\bibinfo {title} {Reporting physisorption data
  for gas/solid systems with special reference to the determination of surface
  area and porosity (recommendations 1984)},\ }\href@noop {} {\bibfield
  {journal} {\bibinfo  {journal} {Pure Appl. Chem.}\ }\textbf {\bibinfo
  {volume} {57}},\ \bibinfo {pages} {603} (\bibinfo {year} {1985})}\BibitemShut
  {NoStop}%
\bibitem [{\citenamefont {Thommes}\ \emph {et~al.}(2015)\citenamefont
  {Thommes}, \citenamefont {Kaneko}, \citenamefont {Neimark}, \citenamefont
  {Olivier}, \citenamefont {Rodriguez-Reinoso}, \citenamefont {Rouquerol},\
  and\ \citenamefont {Sing}}]{IUPAC_2015}%
  \BibitemOpen
  \bibfield  {author} {\bibinfo {author} {\bibfnamefont {M.}~\bibnamefont
  {Thommes}}, \bibinfo {author} {\bibfnamefont {K.}~\bibnamefont {Kaneko}},
  \bibinfo {author} {\bibfnamefont {A.}~\bibnamefont {Neimark}}, \bibinfo
  {author} {\bibfnamefont {J.}~\bibnamefont {Olivier}}, \bibinfo {author}
  {\bibfnamefont {F.}~\bibnamefont {Rodriguez-Reinoso}}, \bibinfo {author}
  {\bibfnamefont {J.}~\bibnamefont {Rouquerol}},\ and\ \bibinfo {author}
  {\bibfnamefont {K.}~\bibnamefont {Sing}},\ }\bibfield  {title} {\bibinfo
  {title} {Physisorption of gases, with special reference to the evaluation of
  surface area and pore size distribution (iupac technical report)},\
  }\href@noop {} {\bibfield  {journal} {\bibinfo  {journal} {Pure Appl. Chem.}\
  }\textbf {\bibinfo {volume} {87}},\ \bibinfo {pages} {1051–1069} (\bibinfo
  {year} {2015})}\BibitemShut {NoStop}%
\bibitem [{\citenamefont {Washburn}(1921)}]{Washburn_1921}%
  \BibitemOpen
  \bibfield  {author} {\bibinfo {author} {\bibfnamefont {E.}~\bibnamefont
  {Washburn}},\ }\bibfield  {title} {\bibinfo {title} {The dynamics of
  capillary flow},\ }\href@noop {} {\bibfield  {journal} {\bibinfo  {journal}
  {Phys. Rev.}\ }\textbf {\bibinfo {volume} {17}},\ \bibinfo {pages}
  {273–283} (\bibinfo {year} {1921})}\BibitemShut {NoStop}%
\bibitem [{\citenamefont {Giesche}(2006)}]{Giesche_2006}%
  \BibitemOpen
  \bibfield  {author} {\bibinfo {author} {\bibfnamefont {H.}~\bibnamefont
  {Giesche}},\ }\bibfield  {title} {\bibinfo {title} {Mercury porosimetry: A
  general (practical) overview},\ }\href@noop {} {\bibfield  {journal}
  {\bibinfo  {journal} {Part. Part. Syst. Charact.}\ }\textbf {\bibinfo
  {volume} {23}},\ \bibinfo {pages} {9} (\bibinfo {year} {2006})}\BibitemShut
  {NoStop}%
\bibitem [{\citenamefont {Cazacliu}(2008)}]{Cazacliu:2008}%
  \BibitemOpen
  \bibfield  {author} {\bibinfo {author} {\bibfnamefont {B.}~\bibnamefont
  {Cazacliu}},\ }\bibfield  {title} {\bibinfo {title} {In-mixer measurements
  for describing mixture evolution during concrete mixing},\ }\href@noop {}
  {\bibfield  {journal} {\bibinfo  {journal} {Cem. Concr. Res.}\ }\textbf
  {\bibinfo {volume} {86}},\ \bibinfo {pages} {1423–1433} (\bibinfo {year}
  {2008})}\BibitemShut {NoStop}%
\bibitem [{\citenamefont {Tomb{\'a}cz}\ \emph {et~al.}(2001)\citenamefont
  {Tomb{\'a}cz}, \citenamefont {Szekeres},\ and\ \citenamefont
  {Klumpp}}]{Tombacz:2001}%
  \BibitemOpen
  \bibfield  {author} {\bibinfo {author} {\bibfnamefont {E.}~\bibnamefont
  {Tomb{\'a}cz}}, \bibinfo {author} {\bibfnamefont {M.}~\bibnamefont
  {Szekeres}},\ and\ \bibinfo {author} {\bibfnamefont {E.}~\bibnamefont
  {Klumpp}},\ }\bibfield  {title} {\bibinfo {title} {Interfacial acid-base
  reactions of aluminum oxide dispersed in aqueous electrolyte solutions. {2.}
  {C}alorimetric study on ionization of surface sites},\ }\href@noop {}
  {\bibfield  {journal} {\bibinfo  {journal} {Langmuir}\ }\textbf {\bibinfo
  {volume} {17}},\ \bibinfo {pages} {1420} (\bibinfo {year}
  {2001})}\BibitemShut {NoStop}%
\bibitem [{\citenamefont {Sun}\ and\ \citenamefont {Berg}(2003)}]{Sun:2003}%
  \BibitemOpen
  \bibfield  {author} {\bibinfo {author} {\bibfnamefont {C.}~\bibnamefont
  {Sun}}\ and\ \bibinfo {author} {\bibfnamefont {J.~C.}\ \bibnamefont {Berg}},\
  }\bibfield  {title} {\bibinfo {title} {A review of the different techniques
  for solid surface acid-base characterization},\ }\href@noop {} {\bibfield
  {journal} {\bibinfo  {journal} {Adv. Colloid Interface Sci.}\ }\textbf
  {\bibinfo {volume} {105}},\ \bibinfo {pages} {151–175} (\bibinfo {year}
  {2003})}\BibitemShut {NoStop}%
\bibitem [{\citenamefont {Cazacliu}\ and\ \citenamefont
  {Roquet}(2009)}]{Cazacliu:2009}%
  \BibitemOpen
  \bibfield  {author} {\bibinfo {author} {\bibfnamefont {B.}~\bibnamefont
  {Cazacliu}}\ and\ \bibinfo {author} {\bibfnamefont {N.}~\bibnamefont
  {Roquet}},\ }\bibfield  {title} {\bibinfo {title} {Concrete mixing kinetics
  by means of power measurement},\ }\href@noop {} {\bibfield  {journal}
  {\bibinfo  {journal} {Cem. Concr. Res.}\ }\textbf {\bibinfo {volume} {39}},\
  \bibinfo {pages} {182} (\bibinfo {year} {2009})}\BibitemShut {NoStop}%
\bibitem [{\citenamefont {Moreno~Juez}\ \emph {et~al.}(2017)\citenamefont
  {Moreno~Juez}, \citenamefont {Artoni},\ and\ \citenamefont
  {Cazacliu}}]{MorenoJuez:2017}%
  \BibitemOpen
  \bibfield  {author} {\bibinfo {author} {\bibfnamefont {J.}~\bibnamefont
  {Moreno~Juez}}, \bibinfo {author} {\bibfnamefont {R.}~\bibnamefont
  {Artoni}},\ and\ \bibinfo {author} {\bibfnamefont {B.}~\bibnamefont
  {Cazacliu}},\ }\bibfield  {title} {\bibinfo {title} {Monitoring of concrete
  mixing evolution using image analysis},\ }\href@noop {} {\bibfield  {journal}
  {\bibinfo  {journal} {Powder Technol.}\ }\textbf {\bibinfo {volume} {305}},\
  \bibinfo {pages} {477–487} (\bibinfo {year} {2017})}\BibitemShut {NoStop}%
\bibitem [{Note1()}]{Note1}%
  \BibitemOpen
  \bibinfo {note} {The number of pores of size $D_p$ can be estimated as the
  ratio of the porous volume associated with the pore diameter $D_p$ in the
  pore size distribution to the volume of either a spherical pore of diameter
  $D_p$ or a cylindrical pore with diameter $D_p$ and typical length 20~nm. We
  checked that both estimates yield the same trends for the evolution of the
  number of pores.}\BibitemShut {Stop}%
\bibitem [{\citenamefont {Alphonse}\ and\ \citenamefont
  {Courty}(2005)}]{Alphonse:2005}%
  \BibitemOpen
  \bibfield  {author} {\bibinfo {author} {\bibfnamefont {P.}~\bibnamefont
  {Alphonse}}\ and\ \bibinfo {author} {\bibfnamefont {M.}~\bibnamefont
  {Courty}},\ }\bibfield  {title} {\bibinfo {title} {Structure and thermal
  behavior of nanocrystalline boehmite},\ }\href@noop {} {\bibfield  {journal}
  {\bibinfo  {journal} {Thermochim. Acta}\ }\textbf {\bibinfo {volume} {425}},\
  \bibinfo {pages} {75–89} (\bibinfo {year} {2005})}\BibitemShut {NoStop}%
\bibitem [{\citenamefont {Rosenberg}\ \emph {et~al.}(1995)\citenamefont
  {Rosenberg}, \citenamefont {Kolenda}, \citenamefont {Szymanski},\ and\
  \citenamefont {Walter}}]{Rosenberg:1995}%
  \BibitemOpen
  \bibfield  {author} {\bibinfo {author} {\bibfnamefont {E.}~\bibnamefont
  {Rosenberg}}, \bibinfo {author} {\bibfnamefont {F.}~\bibnamefont {Kolenda}},
  \bibinfo {author} {\bibfnamefont {R.}~\bibnamefont {Szymanski}},\ and\
  \bibinfo {author} {\bibfnamefont {M.}~\bibnamefont {Walter}},\ }\bibfield
  {title} {\bibinfo {title} {Characterization of alumina paste by
  cryo-microscopy},\ }in\ \href@noop {} {\emph {\bibinfo {booktitle} {Studies
  in surface Science and Catalysis}}},\ Vol.~\bibinfo {volume} {91}\ (\bibinfo
  {publisher} {Elsevier},\ \bibinfo {year} {1995})\ pp.\ \bibinfo {pages}
  {843--850}\BibitemShut {NoStop}%
\bibitem [{\citenamefont {Ito}(1991)}]{Ito:1991}%
  \BibitemOpen
  \bibfield  {author} {\bibinfo {author} {\bibfnamefont {M.}~\bibnamefont
  {Ito}},\ }\bibfield  {title} {\bibinfo {title} {In vitro properties of a
  chitosan-bonded hydroxyapatite bone-filling paste},\ }\href@noop {}
  {\bibfield  {journal} {\bibinfo  {journal} {Biomaterials}\ }\textbf {\bibinfo
  {volume} {12}},\ \bibinfo {pages} {41} (\bibinfo {year} {1991})}\BibitemShut
  {NoStop}%
\bibitem [{\citenamefont {Karouia}\ \emph
  {et~al.}(2013{\natexlab{b}})\citenamefont {Karouia}, \citenamefont
  {Boualleg}, \citenamefont {Digne},\ and\ \citenamefont
  {Alphonse}}]{Karouia:2013}%
  \BibitemOpen
  \bibfield  {author} {\bibinfo {author} {\bibfnamefont {F.}~\bibnamefont
  {Karouia}}, \bibinfo {author} {\bibfnamefont {M.}~\bibnamefont {Boualleg}},
  \bibinfo {author} {\bibfnamefont {M.}~\bibnamefont {Digne}},\ and\ \bibinfo
  {author} {\bibfnamefont {P.}~\bibnamefont {Alphonse}},\ }\bibfield  {title}
  {\bibinfo {title} {Control of the textural properties of nanocrystalline
  boehmite ($\gamma$-{AlOOH}) regarding its peptization ability},\ }\href@noop
  {} {\bibfield  {journal} {\bibinfo  {journal} {Powder Technol.}\ }\textbf
  {\bibinfo {volume} {237}},\ \bibinfo {pages} {602} (\bibinfo {year}
  {2013}{\natexlab{b}})}\BibitemShut {NoStop}%
\bibitem [{\citenamefont {Cassiano-Gaspar}\ \emph {et~al.}(2014)\citenamefont
  {Cassiano-Gaspar}, \citenamefont {Bazer-Bachi}, \citenamefont {Chevalier},
  \citenamefont {L{\'e}colier}, \citenamefont {Jorand},\ and\ \citenamefont
  {Rouleau}}]{Cassiano:2014}%
  \BibitemOpen
  \bibfield  {author} {\bibinfo {author} {\bibfnamefont {S.}~\bibnamefont
  {Cassiano-Gaspar}}, \bibinfo {author} {\bibfnamefont {D.}~\bibnamefont
  {Bazer-Bachi}}, \bibinfo {author} {\bibfnamefont {J.}~\bibnamefont
  {Chevalier}}, \bibinfo {author} {\bibfnamefont {E.}~\bibnamefont
  {L{\'e}colier}}, \bibinfo {author} {\bibfnamefont {Y.}~\bibnamefont
  {Jorand}},\ and\ \bibinfo {author} {\bibfnamefont {L.}~\bibnamefont
  {Rouleau}},\ }\bibfield  {title} {\bibinfo {title} {Novel extrudates based on
  the multiscale packing of alumina particles and boehmite or aluminophosphate
  binders},\ }\href@noop {} {\bibfield  {journal} {\bibinfo  {journal} {Powder
  Technol.}\ }\textbf {\bibinfo {volume} {255}},\ \bibinfo {pages} {74}
  (\bibinfo {year} {2014})}\BibitemShut {NoStop}%
\bibitem [{\citenamefont {Thompson}\ \emph {et~al.}(2012)\citenamefont
  {Thompson}, \citenamefont {Weatherley}, \citenamefont {Pukadyil},\ and\
  \citenamefont {Sheskey}}]{Thompson:2012}%
  \BibitemOpen
  \bibfield  {author} {\bibinfo {author} {\bibfnamefont {M.}~\bibnamefont
  {Thompson}}, \bibinfo {author} {\bibfnamefont {S.}~\bibnamefont
  {Weatherley}}, \bibinfo {author} {\bibfnamefont {R.}~\bibnamefont
  {Pukadyil}},\ and\ \bibinfo {author} {\bibfnamefont {P.}~\bibnamefont
  {Sheskey}},\ }\bibfield  {title} {\bibinfo {title} {Foam granulation: new
  developments in pharmaceutical solid oral dosage forms using twin screw
  extrusion machinery},\ }\href@noop {} {\bibfield  {journal} {\bibinfo
  {journal} {Drug Dev. Ind. Pharm.}\ }\textbf {\bibinfo {volume} {38}},\
  \bibinfo {pages} {771} (\bibinfo {year} {2012})}\BibitemShut {NoStop}%
\bibitem [{\citenamefont {Li}\ \emph {et~al.}(2014)\citenamefont {Li},
  \citenamefont {Thompson},\ and\ \citenamefont {O’donnell}}]{Li:2014}%
  \BibitemOpen
  \bibfield  {author} {\bibinfo {author} {\bibfnamefont {H.}~\bibnamefont
  {Li}}, \bibinfo {author} {\bibfnamefont {M.}~\bibnamefont {Thompson}},\ and\
  \bibinfo {author} {\bibfnamefont {K.}~\bibnamefont {O’donnell}},\
  }\bibfield  {title} {\bibinfo {title} {Understanding wet granulation in the
  kneading block of twin screw extruders},\ }\href@noop {} {\bibfield
  {journal} {\bibinfo  {journal} {Chem. Eng. Sci.}\ }\textbf {\bibinfo {volume}
  {113}},\ \bibinfo {pages} {11} (\bibinfo {year} {2014})}\BibitemShut
  {NoStop}%
\bibitem [{\citenamefont {Iveson}\ \emph {et~al.}(2001)\citenamefont {Iveson},
  \citenamefont {Litster}, \citenamefont {Hapgood},\ and\ \citenamefont
  {Ennis}}]{Iveson_2001}%
  \BibitemOpen
  \bibfield  {author} {\bibinfo {author} {\bibfnamefont {S.~M.}\ \bibnamefont
  {Iveson}}, \bibinfo {author} {\bibfnamefont {J.~D.}\ \bibnamefont {Litster}},
  \bibinfo {author} {\bibfnamefont {K.}~\bibnamefont {Hapgood}},\ and\ \bibinfo
  {author} {\bibfnamefont {B.~J.}\ \bibnamefont {Ennis}},\ }\bibfield  {title}
  {\bibinfo {title} {Nucleation, growth and breakage phenomena in agitated wet
  granulation processes: a review},\ }\href@noop {} {\bibfield  {journal}
  {\bibinfo  {journal} {Powder Technol.}\ }\textbf {\bibinfo {volume} {117}},\
  \bibinfo {pages} {3} (\bibinfo {year} {2001})}\BibitemShut {NoStop}%
\bibitem [{\citenamefont {Collet}\ \emph {et~al.}(2011)\citenamefont {Collet},
  \citenamefont {Oulahna}, \citenamefont {{De Ryck}},\ and\ \citenamefont
  {Jezequel}}]{Collet_2011}%
  \BibitemOpen
  \bibfield  {author} {\bibinfo {author} {\bibfnamefont {R.}~\bibnamefont
  {Collet}}, \bibinfo {author} {\bibfnamefont {D.}~\bibnamefont {Oulahna}},
  \bibinfo {author} {\bibfnamefont {A.}~\bibnamefont {{De Ryck}}},\ and\
  \bibinfo {author} {\bibfnamefont {M.}~\bibnamefont {Jezequel}, \bibfnamefont
  {P.~H.and~Martin}},\ }\bibfield  {title} {\bibinfo {title} {Mixing of a wet
  granular medium: Influence of the liquid addition method},\ }\href@noop {}
  {\bibfield  {journal} {\bibinfo  {journal} {Powder Technol.}\ }\textbf
  {\bibinfo {volume} {208}},\ \bibinfo {pages} {367–371} (\bibinfo {year}
  {2011})}\BibitemShut {NoStop}%
\bibitem [{\citenamefont {Rondet}\ \emph
  {et~al.}(2009{\natexlab{b}})\citenamefont {Rondet}, \citenamefont
  {Rundgsiyopas}, \citenamefont {Ruiz}, \citenamefont {Delalonde},\ and\
  \citenamefont {Desfours}}]{Rondet:2009}%
  \BibitemOpen
  \bibfield  {author} {\bibinfo {author} {\bibfnamefont {E.}~\bibnamefont
  {Rondet}}, \bibinfo {author} {\bibfnamefont {M.}~\bibnamefont
  {Rundgsiyopas}}, \bibinfo {author} {\bibfnamefont {T.}~\bibnamefont {Ruiz}},
  \bibinfo {author} {\bibfnamefont {M.}~\bibnamefont {Delalonde}},\ and\
  \bibinfo {author} {\bibfnamefont {J.-P.}\ \bibnamefont {Desfours}},\
  }\bibfield  {title} {\bibinfo {title} {Hydrotextural description of an
  unsaturated humid granular media: application for kneading, packing and
  drying operations},\ }\href@noop {} {\bibfield  {journal} {\bibinfo
  {journal} {KONA Powder Part. J.}\ }\textbf {\bibinfo {volume} {27}},\
  \bibinfo {pages} {174} (\bibinfo {year} {2009}{\natexlab{b}})}\BibitemShut
  {NoStop}%
\bibitem [{\citenamefont {Romanova}\ \emph {et~al.}(2022)\citenamefont
  {Romanova}, \citenamefont {Nadeina}, \citenamefont {Danilova}, \citenamefont
  {Danilevich}, \citenamefont {Pakharukova}, \citenamefont {Gabrienko},
  \citenamefont {Glazneva}, \citenamefont {Gerasimov}, \citenamefont
  {Prosvirin}, \citenamefont {Vatutina} \emph {et~al.}}]{Romanova:2022}%
  \BibitemOpen
  \bibfield  {author} {\bibinfo {author} {\bibfnamefont {T.}~\bibnamefont
  {Romanova}}, \bibinfo {author} {\bibfnamefont {K.}~\bibnamefont {Nadeina}},
  \bibinfo {author} {\bibfnamefont {I.}~\bibnamefont {Danilova}}, \bibinfo
  {author} {\bibfnamefont {V.}~\bibnamefont {Danilevich}}, \bibinfo {author}
  {\bibfnamefont {V.}~\bibnamefont {Pakharukova}}, \bibinfo {author}
  {\bibfnamefont {A.}~\bibnamefont {Gabrienko}}, \bibinfo {author}
  {\bibfnamefont {T.}~\bibnamefont {Glazneva}}, \bibinfo {author}
  {\bibfnamefont {E.~Y.}\ \bibnamefont {Gerasimov}}, \bibinfo {author}
  {\bibfnamefont {I.}~\bibnamefont {Prosvirin}}, \bibinfo {author}
  {\bibfnamefont {Y.~V.}\ \bibnamefont {Vatutina}}, \emph {et~al.},\ }\bibfield
   {title} {\bibinfo {title} {Modification of hdt catalysts of fcc feedstock by
  adding silica to the kneading paste of alumina support: Advantages and
  disadvantages},\ }\href@noop {} {\bibfield  {journal} {\bibinfo  {journal}
  {Fuel}\ }\textbf {\bibinfo {volume} {324}},\ \bibinfo {pages} {124555}
  (\bibinfo {year} {2022})}\BibitemShut {NoStop}%
\bibitem [{\citenamefont {Caponio}\ \emph {et~al.}(1999)\citenamefont
  {Caponio}, \citenamefont {Alloggio},\ and\ \citenamefont
  {Gomes}}]{Caponio:1999}%
  \BibitemOpen
  \bibfield  {author} {\bibinfo {author} {\bibfnamefont {F.}~\bibnamefont
  {Caponio}}, \bibinfo {author} {\bibfnamefont {V.}~\bibnamefont {Alloggio}},\
  and\ \bibinfo {author} {\bibfnamefont {T.}~\bibnamefont {Gomes}},\ }\bibfield
   {title} {\bibinfo {title} {Phenolic compounds of virgin olive oil: influence
  of paste preparation techniques},\ }\href@noop {} {\bibfield  {journal}
  {\bibinfo  {journal} {Food Chem.}\ }\textbf {\bibinfo {volume} {64}},\
  \bibinfo {pages} {203} (\bibinfo {year} {1999})}\BibitemShut {NoStop}%
\bibitem [{\citenamefont {Lombois-Burger}\ \emph {et~al.}(2006)\citenamefont
  {Lombois-Burger}, \citenamefont {Colombet}, \citenamefont {Halary},\ and\
  \citenamefont {Van~Damme}}]{Lombois:2006}%
  \BibitemOpen
  \bibfield  {author} {\bibinfo {author} {\bibfnamefont {H.}~\bibnamefont
  {Lombois-Burger}}, \bibinfo {author} {\bibfnamefont {P.}~\bibnamefont
  {Colombet}}, \bibinfo {author} {\bibfnamefont {J.~L.}\ \bibnamefont
  {Halary}},\ and\ \bibinfo {author} {\bibfnamefont {H.}~\bibnamefont
  {Van~Damme}},\ }\bibfield  {title} {\bibinfo {title} {Kneading and extrusion
  of dense polymer--cement pastes},\ }\href@noop {} {\bibfield  {journal}
  {\bibinfo  {journal} {Cem. Concr. Res.}\ }\textbf {\bibinfo {volume} {36}},\
  \bibinfo {pages} {2086} (\bibinfo {year} {2006})}\BibitemShut {NoStop}%
\bibitem [{\citenamefont {Kuroda}\ and\ \citenamefont
  {Scott}(2002)}]{Kuroda:2002}%
  \BibitemOpen
  \bibfield  {author} {\bibinfo {author} {\bibfnamefont {M.~M.}\ \bibnamefont
  {Kuroda}}\ and\ \bibinfo {author} {\bibfnamefont {C.~E.}\ \bibnamefont
  {Scott}},\ }\bibfield  {title} {\bibinfo {title} {Blade geometry effects on
  initial dispersion of chopped glass fibers},\ }\href@noop {} {\bibfield
  {journal} {\bibinfo  {journal} {Polym. Compos.}\ }\textbf {\bibinfo {volume}
  {23}},\ \bibinfo {pages} {828} (\bibinfo {year} {2002})}\BibitemShut
  {NoStop}%
\bibitem [{\citenamefont {Wu}\ and\ \citenamefont {Wei}(2000)}]{Wu_1999}%
  \BibitemOpen
  \bibfield  {author} {\bibinfo {author} {\bibfnamefont {R.-Y.}\ \bibnamefont
  {Wu}}\ and\ \bibinfo {author} {\bibfnamefont {W.-C.~J.}\ \bibnamefont
  {Wei}},\ }\bibfield  {title} {\bibinfo {title} {Torque evolution and effects
  on alumina feedstocks prepared by various kneading sequences},\ }\href@noop
  {} {\bibfield  {journal} {\bibinfo  {journal} {J. Eur. Ceram.}\ }\textbf
  {\bibinfo {volume} {20}},\ \bibinfo {pages} {67} (\bibinfo {year}
  {2000})}\BibitemShut {NoStop}%
\bibitem [{\citenamefont {Salomão}(2018)}]{Salomao:2018}%
  \BibitemOpen
  \bibfield  {author} {\bibinfo {author} {\bibfnamefont {R.}~\bibnamefont
  {Salomão}},\ }\bibfield  {title} {\bibinfo {title} {Porogenic behavior of
  water in high-alumina castable structures},\ }\href@noop {} {\bibfield
  {journal} {\bibinfo  {journal} {Adv. Mater. Sci. Eng.}\ }\textbf {\bibinfo
  {volume} {2018}} (\bibinfo {year} {2018})}\BibitemShut {NoStop}%
\end{thebibliography}

\providecommand{\noopsort}[1]{}\providecommand{\singleletter}[1]{#1}%

\end{document}